\documentclass[12pt,draftclsnofoot,onecolumn,compsoc]{IEEEtran}

\ifCLASSOPTIONcompsoc
\usepackage[nocompress]{cite}
\else
\usepackage{cite}
\fi

\usepackage{dsfont}
\usepackage{amssymb}
\usepackage{subfig}
\usepackage{graphicx, amssymb, amsmath}
\usepackage{extarrows}
\usepackage{algorithm} %format of the algorithm
\usepackage[noend]{algpseudocode}
\usepackage{multirow} %multirow for format of table
\usepackage{color}
\usepackage{amsmath}
\usepackage{url}
\usepackage{ulem}
\emergencystretch=\maxdimen
\IEEEoverridecommandlockouts 
\begin{document}

%\title{Joint Cache Update and Content Delivery: Providing Fresh Content with Low-Latency in Information-Centric Vehicular Networks}

\title{Low-Latency and Fresh Content Provision in Information-Centric Vehicular Networks}

\author{Shan~Zhang,~\IEEEmembership{Member,~IEEE,}
	Junjie~Li,%~\IEEEmembership{Student~Member,~IEEE,}
	Hongbin~Luo,~\IEEEmembership{Member,~IEEE,}
	Jie~Gao,~~\IEEEmembership{Member,~IEEE,}
	Lian~Zhao,~\IEEEmembership{Senior~Member,~IEEE,}
	and~Xuemin~(Sherman)~Shen,~\IEEEmembership{Fellow,~IEEE}% <-this % stops a space
	\IEEEcompsocitemizethanks{\IEEEcompsocthanksitem Shan~Zhang, Junjie Li, and Hongbin~Luo are with Beijing Key Laboratory of Computer Networks, the School of Computer Science and Engineering, Beihang University, Beijing, China.\protect\\
		E-mail: \{zhangshan18, ljj0618, luohb\}@buaa.edu.cn
	\IEEEcompsocthanksitem Jie Gao and Xuemin~(Sherman)~Shen are with the Department of Electrical and Computer Engineering, University of Waterloo, 200 University Avenue West, Waterloo, Ontario, Canada, N2L 3G1.\protect\\
		E-mail: \{jie.gao, sshen\}@uwaterloo.ca
	\IEEEcompsocthanksitem Lian Zhao is with the Department of Electrical, Computer, and Biomedical Engineering, Ryerson University, Toronto, ON, Canada.\protect\\
		E-mail: l5zhao@ryerson.ca}% <-this % stops an unwanted space
	\thanks{This work is supported in part by the Nature Science Foundation of China No. 61801011, 91638204, in part by the Fundamental Research Funds for the Central Universities, China, under Grant No. YWF-18-BJ-J-61.}% <-this % stops a space
	%\thanks{Part of this work has been presented in the International Conference on Wireless Communications and Signal Processing (WCSP) 2018 \cite{mine_AoI_WCSP}.}
}

%%%\maketitle
%\markboth{IEEE Transactions on Mobile Computing,~Vol.~~, No.~~, Nov.~2019}%
%{Zhang \MakeLowercase{\textit{et al.}}: Low-Latency and Fresh Content Provision in Information-Centric Vehicular Networks}

\IEEEtitleabstractindextext{%
\begin{abstract}
	
	In this paper, the content service provision of information-centric vehicular networks (ICVNs) is investigated from the aspect of mobile edge caching, considering the dynamic driving-related context information.
	To provide up-to-date information with low latency, two schemes are designed for cache update and content delivery at the roadside units (RSUs).
	The roadside unit centric (RSUC) scheme decouples cache update and content delivery through bandwidth splitting, where the cached content items are updated regularly in a round-robin manner.
	The request adaptive (ReA) scheme updates the cached content items upon user requests with certain probabilities.
	The performance of both proposed schemes are analyzed, whereby the average age of information (AoI) and service latency are derived in closed forms.
	Surprisingly, the AoI-latency trade-off does not always exist, and frequent cache update can degrade both performances.
	Thus, the RSUC and ReA schemes are further optimized to balance the AoI and latency.
	Extensive simulations are conducted on SUMO and OMNeT++ simulators, and the results show that the proposed schemes can reduce service latency by up to 80\% while guaranteeing content freshness in heavily loaded ICVNs.
	
\end{abstract}

}

\maketitle
\IEEEdisplaynontitleabstractindextext

\IEEEpeerreviewmaketitle

\IEEEraisesectionheading{\section{Introduction}\label{sec:introduction}}

\IEEEPARstart{V}{ehicular} communications are featured by high mobility, short-lived link connectivity, and rapidly changing network topology, posing great challenges to guarantee network efficiency and reliability \cite{Shah18_VANET_5G}.
In this context, Information Centric Vehicular Networking (ICVN) is emerging as a promising paradigm, which supports the receiver-driven content retrieval without requiring the conventional sender-driven end-to-end connectivity \cite{Grewe18_survey_ICN_VN_arch}.
Furthermore, ICVN can utilize the storage resources of roadside units (RSUs) or vehicles to realize mobile edge caching, bringing three-fold benefits of reduced end-to-end latency, high transmission efficiency, and enhanced system reliability\cite{Poularakis17_cache_mobility_TMC}. % 展开说嘛？	Firstly, the RSUs can serve vehicles directly through on-hop without fetching the information from sources, reducing the end-to-end service latency.  Secondly, more spectrum can be used for content delivery to improve network capacity, which is especially beneficial for heavily-loaded networks. Thirdly, in case of deep fading channels, the RSU can still provide service timely regardless of transmission failure from sources, enhancing system reliability.
However, in practice, mobile edge caching usually faces the contradictory of limited storage resource and massive data, raising fundamental issues including where to cache, what to cache, and how to update \cite{Gao19_probabilistic_cache_1,mine_cache_TMC,Jiang17_SBS_UE_caching_TMC,Qu19_coop_cache_TMC}.
Specifically, cache update is critical in the dynamic ICVN environment.
On the one hand, the popularity of content items may change with time, since new items keep generating (such as news and entertainment information) \cite{Zhao17_ICVN_popul_replace_Access,Hajiakhondi18_cache_replace_TMC}.
On the other hand, there exists extensive driving-related context information which may vary with time, such as the status of traffic lights, availability of parking lot, promotion of stores \cite{mine_auto_drive_mag}.
For the former case, the newly generated popular content items should be cached and replace the unpopular ones, to keep high hit rate at the network edge.
The latter case requires to download new versions of the cached items from time to time to guarantee the content effectiveness.
Extensive efforts have been devoted on cache update considering the content popularity variation, including popularity prediction and content replacement algorithm design \cite{YangPeng19_cache_TMM,Gao19_probabilistic_cache_2}.
However, the dynamic content variations of cached items has been seldom considered.

In this work, we jointly design cache caching and content delivery of RSUs in ICVNs, considering the dynamic context information services.
The objective is to provide fresh information to vehicles rapidly on demand.
A typical ICVN scenario is considered, which consists of publishers generating content items, cache-enabled RSUs providing content services, and moving vehicles randomly raising content requests.
As the context information may change with the driving environment, the publishers continuously generate new versions of content items to reflect the real-time status.
Meanwhile, RSUs update the cache timely to deliver fresh content to vehicles.
Notice that cache update will consume additional bandwidth resources, which may degrade the content delivery efficiency and introduce longer service latency.
Thus, cache update and content delivery should be jointly designed to balance content freshness and service latency.
In this regard, two schemes are proposed, i.e., RoadSide Unit Centric (RSUC) scheme and Request Adaptive (ReA) scheme.
The RSUC scheme decouples content update and content delivery by splitting the bandwidth.
All cached items are updated in a round-robin manner, while the RSU will serve the requests with the current version of cached content items in a First-Come-First-Serve (FCFS) policy.
The ReA scheme updates cached items along with content delivery.
Specifically, a request triggers cache update with a certain probability, wherein the RSU will fetch the new version of the request item before delivery.
Two substantially different performance metrics, i.e., Age of Information (AoI) and service latency are adopted to analyze the proposed schemes.
AoI, i.e., the time elapsed since the generation of the content, depicts the freshness of the content.
Instead, service latency characterizes the response speed of the RSU, i.e., the time needed for a vehicle to receive the content after raising a request.
By applying stochastic process and queueing theory, the average AoI and service latency are derived in closed form under both schemes.
Analytical results show that the content freshness and service latency do not always show a trade-off, and the frequent cache update may increase the average AoI and latency simultaneously.
Therefore, we further optimize the RSUC and ReA schemes to balance the AoI and latency.
Extensive simulations have been conducted using the Monte Carlo method on the OMNeT++ system to validate the theoretical analysis and evaluate the performance of the two schemes under different scenarios and system parameters.
%The main conclusions include (1) both RSUC and ReA schemes can balance AoI and latency requirements by adjusting the operational parameters; (2) the RSUC scheme outperforms the ReA scheme when the average AoI requirement is higher than a certain threshold, and otherwise the ReA scheme is more advantageous; (3) the ReA scheme is more advantageous in case of heavier traffic load or larger number of content items; (4) the RSUC scheme is effective to deal with uniform interests on content items, while ReA scheme can better adapt to the case of concentrated requests.
In addition, a real-trace system level simulation is conducted by using SUMO and OMNeT++ simulation platforms.
The simulation results show that the ReA and RSUC schemes can reduce service latency by 65\% and 80\% compared with the conventional scheme.

The main contributions of this work are as follows:
\begin{itemize}
	\item The joint design of cache update and content delivery is studied under the ICVN architecture considering the content dynamics, where content freshness and service latency are both guaranteed to enhance the quality of experience;
	\item The average AoI and service latency are derived under the proposed schemes, whereby the freshness-latency interplay is quantified through theoretical analysis;
	\item The proposed schemes are further optimized for AoI-latency balance on the demand of applications.
\end{itemize}

The remaining of this paper is organized as follows. 
Section~\ref{sec_review} reviews related works.
Section~\ref{sec_system_model} introduces the architecture of ICVN and proposes cache update and content delivery schemes.
The performances of the proposed schemes are analyzed in Section \ref{sec_proposed_analysis}, whereby the interplay between AoI and latency performances is studied in Section~\ref{sec_tradeoff}. 
Simulation results are provided in Section~\ref{sec_simulation}, followed by conclusions in Section~\ref{sec_conclusions}.

%%%%%%%%%%%%%%%%%%%%%%%%%%%%%%%%%%%%%%%%%%%%%%%%%%%%%%%%%%%%%%%%%%%%%%%%%%%%%%%%%%%%%%%%%%%%%%%%%%%
\section{Literature Review}
\label{sec_review}

	This section provides a comprehensive review on the design of ICVN architecture, in-network caching, AoI and timely content services to highlight the novelty of this work.
	
	\subsection{Vehicular Information Centric Network Architecture}
	
	Extensive efforts have been devoted on ICVN architecture design. 
	Classical Information Centric Networking (ICN) architectures, such as Content Centric Networking (CCN), Named Data Networking (NDN), and MobilityFirst, have been enhanced to adapt to vehicular communication features like high mobility and wireless link uncertainty \cite{Ortega18_CCN_VN_mag, Kalogeiton18_NDN_VANET, Baid13_MobFirst_VN_arch_conf}.
	Grewe \textit{et. al} have shown that the CCN/NDN based approaches can satisfy most of the vehicular communication requirements like naming, data dissemination, safety and security \cite{Grewe18_survey_ICN_VN_arch}.
	Furthermore, Grassi \textit{et. al} have built an NDN-based vehicular communication prototype \cite{Grassi14_NDN_VN_prototype_inforcom}.
	Content dissemination is a critical management issue due to the vehicular mobility and link unreliability \cite{He17_TMC_carry_forward_VANET}.
	A location-based content forwarding scheme has been proposed to allow vehicles fetch data from multiple potential carriers for reliability, which is more robust to vehicle topology variations and link disruptions \cite{Grassi15_NDN_VN_georaph_distri_conf}.
	Boukerche \textit{et. al} have suggested to use neighboring vehicles with higher link reliability, to improve content delivery rate and reduce duplicated network transmissions \cite{Boukerche17_ICVN_distri_link_reliability_MASS}.
	Furthermore, a RSU-centric data dissemination algorithm has been proposed to ensure timely and efficient delivery of safety-related content \cite{Chen17_ICVN_safety_infor_RSU_conf}.	
	
	\subsection{In-network Caching in Vehicular Networks}
	
	In-network caching is one of the key design issues for ICN architectures, and has been extensively studied in the static Internet environments \cite{Zhang15_ICN_cache_survey}.
	However, this problem needs to be revisited for the ICVN considering the short-lived connections, time-varying topology, dynamic context information and applications \cite{Modesto17_ICVN_caching_summary_ICC}.
	%To this end, three fundamental issues are studied, i.e., where to cache, what to cache, and how to update.
	The very recent works have started exploration in this area, including RSU caching \cite{mine_VeCache_COMMAG,Mauri17_ICVN_RSU_core_caching_TVT} vehicle caching \cite{mine_JSAC_VNET,Li18_ICVN_content_distri_crowds_Access,Yao18_vehicle_cache_mobility_TVT,Zhao17_ICVN_popul_replace_Access}.
	
	Caching at RSUs can address the backhaul congestion issues, which is beneficial for both network deployment and service quality enhancement \cite{mine_VeCache_COMMAG}.
	%Simulation results have shown that proactively caching at RSUs can help to reduce content delivery latency, compared with reactive caching \cite{Hasan18_caching_VICN_performance_ICOIN}.
	Efficient RSU content placement algorithm has been proposed to maximize the content retrieval probability by solving an integer linear programming problem, with the cache size optimized \cite{Mauri17_ICVN_RSU_core_caching_TVT}.
	The results have shown that RSU caching is more effective compared with the core-network caching, especially in case of low RSU density or high vehicle density.	
	In addition to RSU caching, the rich storage resources of vehicles can be also utilized.
	Location-based content items can be pushed to the cache of vehicles prior to requests based on the driving trajectory, which can better support vehicle mobility \cite{mine_JSAC_VNET}.
	Cache-enabled vehicles can also form groups for efficient content delivery in a crowdsourcing manner \cite{Li18_ICVN_content_distri_crowds_Access}.
	Specifically, it is more efficient to cache contents on the vehicles which may stay in hot regions for longer time or have more social connectivities \cite{Yao18_vehicle_cache_mobility_TVT}. %Khan16_ICVN_social_vehicle_rank_mag
	The very few works have studied cache update based on the dynamics of content popularity  \cite{Zhao17_ICVN_popul_replace_Access}, whereas the dynamic variation of content information has been ignored.

	\subsection{Age of Information and Timely Service}
	
	The concept of AoI was firstly introduced to capture the requirement of time-critical vehicular safety applications \cite{Kaul11_AOI_concept_conf}.
	In specific, AoI is defined as the time elapsed since the generation of the corresponding content, which quantifies the freshness of knowledge we have about a remote time-varying source \cite{Kosta17_AoI_survey}.
	%As a metric of timeliness, AoI is substantially different from service latency or delay \cite{Kaul12_AoI_basic_update_infocom}.
	Recent works have adopted this new concept to evaluate the performance of dynamic systems, and accordingly new system scheduling and management methods have been proposed.	
	The AoI performance has been analyzed theoretically by applying queueing models in case of single source and multiple sources under different service mode \cite{Kaul12_AoI_basic_update_infocom,Najm16_AoI_performance_gamma_ISIT,Yates18_TIT}.
	Kam \textit{et. al} have further utilized multi-server queueing model to reflect service priority and multi-path transmission diversity \cite{Kam16_AoI_multi_server_TIT}.
	%The theoretical results have shown that the average AoI is a non-monotone convex function with respect to the content update frequency, which can be applied for efficient content update \cite{Kaul12_AoI_basic_update_infocom,Kam16_AoI_multi_server_TIT}.
	AoI-optimal content update schemes have been proposed for single-source and multi-source heterogeneous systems \cite{Costa16_AoI_management_single_source_TIT,Stamatakis18_AoI_management_heterogeneous_source}.
	Furthermore, AoI has been adopted for the metric of cloud game applications to achieve timely video frame transmissions \cite{Yates17_AoI_cloud_game_INFOCOM}.
	
	The very recent works have implemented the AoI in mobile edge caching \cite{Kam17_AoI_cache_fresh_popular_ISIT,Yates17_AoI_cache_update_17,mine_AoI_WCSP}.
	Considering that the popularity of a content item may fade with time, the authors have built a model of request rate based on both historical request rate and content AoI, whereby the optimal content update policy has been devised \cite{Kam17_AoI_cache_fresh_popular_ISIT}.
	However, the dynamic variation of content is ignored.
	\cite{Yates17_AoI_cache_update_17} and \cite{mine_AoI_WCSP} have considered the content dynamics in mobile edge caching, which are the most related to this work.
	Yates \textit{et. al} have studied the content update problem for a local cache system in \cite{Yates17_AoI_cache_update_17} to minimized the average AoI of cached content items.
	An optimization problem is formulated and solved in an approximated way with relaxation.
	The main differences between this work and \cite{Yates17_AoI_cache_update_17} are three-fold:
	(1) we aim at minimizing the AoI of user-received content instead of local-cached content, (2) the content caching and delivery are jointly optimized, and (3) the service latency is also considered in addition to AoI to provide better quality of experience.
	A cache-assisted lazy update and delivery (CALUD) scheme has been proposed in our preliminary work \cite{mine_AoI_WCSP}, which was the first to study the interplay between latency and content freshness performances from the edge caching aspect. 
	Nevertheless, the CALUD scheme is limited to the ideal wireless channel assumption, and each source utilizes dedicated bandwidth in a low-efficient way.

%%%%%%%%%%%%%%%%%%%%%%%%%%%%%%%%%%%%%%%%%%%%%%%%%%%%%%%%%%%%%%%%%%%%%%%%%%%%%%%%%%%%%%%%%%%%%%%%%%%
\section{Information-Centric Vehicular Network}
\label{sec_system_model}

	% To do
	% add a notation table
	
	\begin{figure}[!t]
		\centering
		\includegraphics[width=3.2in]{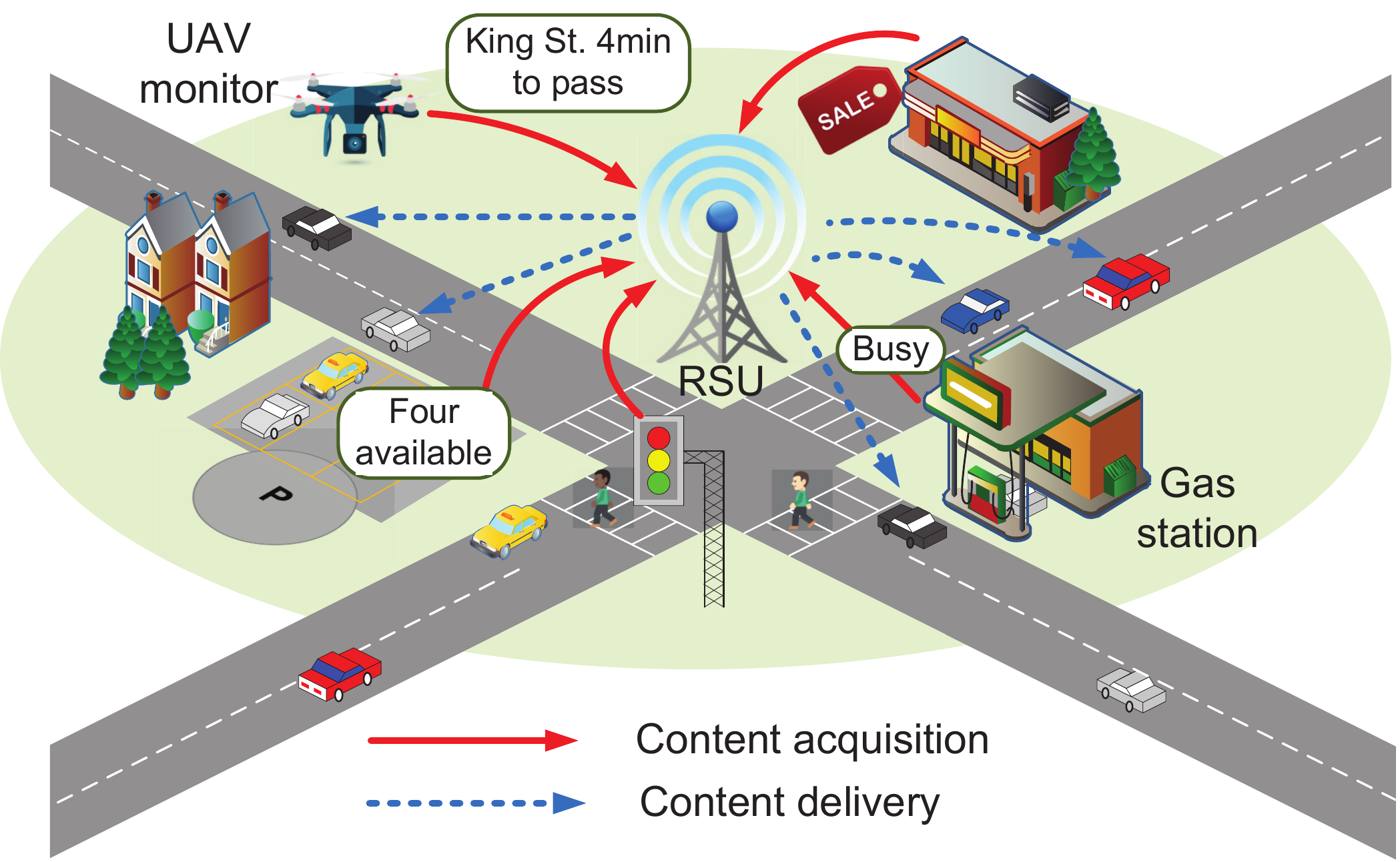}
		\caption{Information-centric vehicular networks.}
		\label{fig_scenario}
	\end{figure}

	%% network scenario
	%\subsection{Network Architecture}
	
	% scenario
	A typical information-centric vehicular network is illustrated in Fig.~\ref{fig_scenario}. 
	We focus on the context information, which is driving-related and helps vehicles to get awareness of surrounding environment.
	Examples can be the occupancy of nearby parking lots, the availability of gas stations, the road congestion status, and the promotion information of shopping malls. 
	The context information is collected by distributed sensors or systems, and then published as content items by corresponding \textit{publishers}.
	Denote by $\mathcal{S}=\{1,2,...,S\}$ the set of content items generated by all publishers, where $S=|\mathcal{S}|$ and each content item has the same size $L$.
	Vehicles raise content requests randomly, and the request arrival process for the content item $s$ is modeled as a Poisson process of rate $\lambda_s$.
	The RSU works as a sink node, which caches all generated content items and makes delivery on the demand of vehicles.
	Denote by $B$ the available bandwidth of the RSU.	
	We propose two service schemes, and the conventional freshness-first scheme is adopted as a baseline.

	\subsection{Conventional Scheme}
	
\begin{figure}[!t]
		\centering
		\includegraphics[width=2.2in]{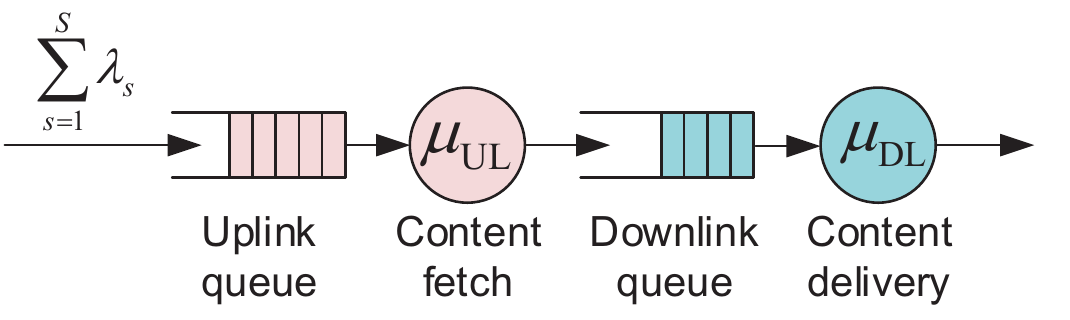}
		\caption{Content service process of the conventional method.}
		\label{fig_queue_conventional}
	\end{figure}

	According to the conventional freshness-first scheme, all the vehicle requests will be served through two-hop transmission.
	In particular, when a vehicle raises a request, the RSU will firstly fetch the content from the corresponding publisher through the uplink channel, and then deliver the content to the vehicle through the downlink channel.
	The key idea is to update the RSU cache frequently, in order to avoid content staleness.
	The uplink and downlink channels share the bandwidth orthogonally, with a bandwidth splitting ratio $\beta$.
	Then, the service process of the RSU can be modeled as a tandem queueing system with two servers, as shown in Fig.~\ref{fig_queue_conventional}.
	The uplink and downlink average transmission rates are given by
	\begin{equation}
	\label{eq_rate}
	\begin{split}
	\mu_\mathrm{UL} & = \frac{\beta B}{L} \log_2 \left( 1 + \gamma_\mathrm{UL}\right) \triangleq \beta R_\mathrm{UL},\\
	\mu_\mathrm{DL} & = \frac{(1-\beta) B}{L} \log_2 \left( 1 + \gamma_\mathrm{DL}\right) \triangleq (1-\beta) R_\mathrm{DL},
	\end{split}
	\end{equation}
	respectively, where $\gamma_\mathrm{UL}$ and $\gamma_\mathrm{DL}$ are the average received signal to interference and noise ratio (SINR) of the uplink and downlink, $R_\mathrm{UL}$ and $R_\mathrm{DL}$ are the normalized uplink and downlink service rates with full bandwidth allocation.

	\subsection{RSU-Centric Scheme}
	
\begin{figure}[!t]
		\centering
		\includegraphics[width=2.8in]{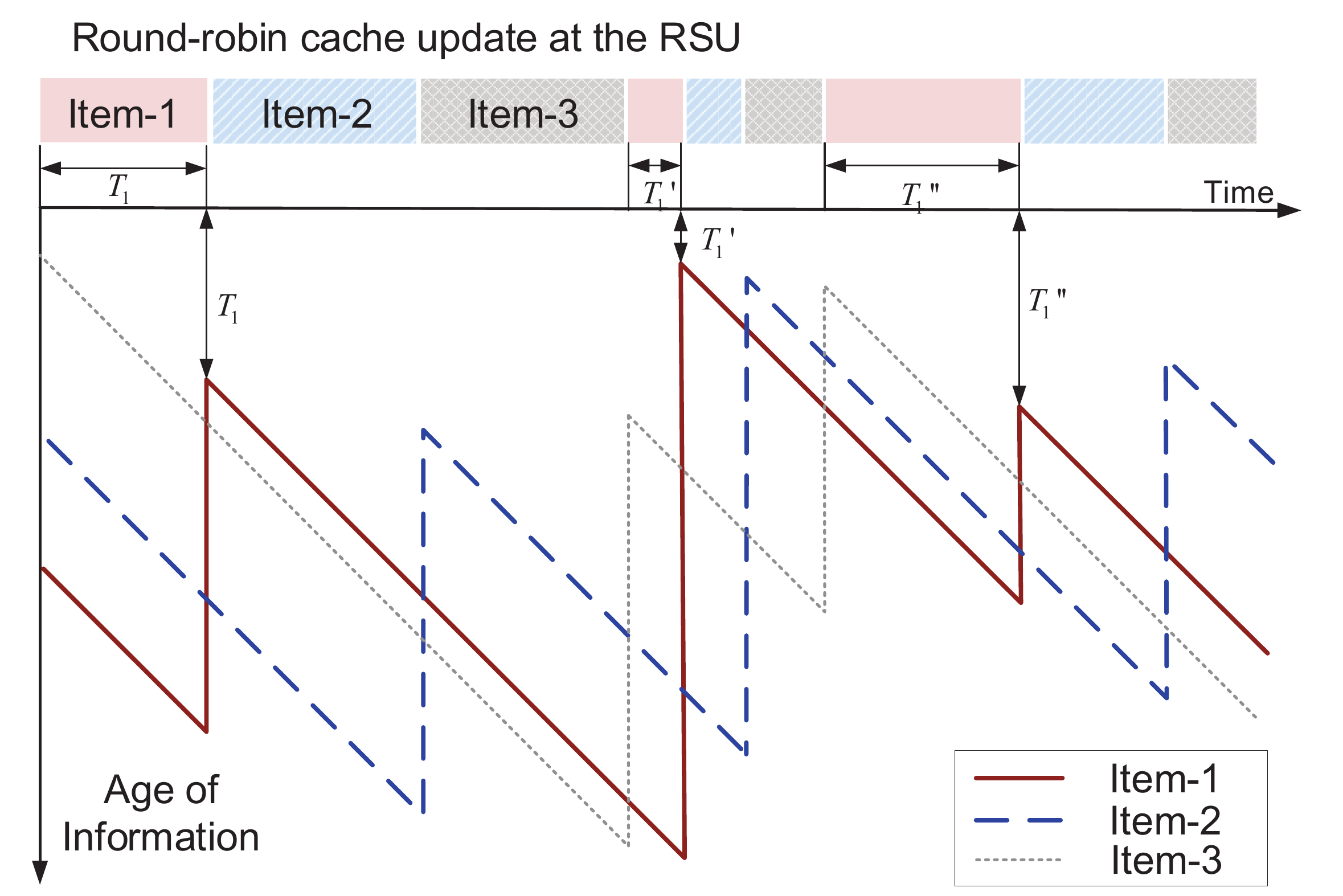}
		\caption{Illustration of AoI variation for the RSU cached content items under the RSUC scheme.}
		\label{fig_AoI}
	\end{figure}

	The RSUC scheme decouples cache content fetching and delivery by utilizing the storage modules.
	In specific, the generated content items are cached at the RSU, and updated in a round-robin manner through uplink transmissions.
	When a vehicle raises a request, the RSU will directly send the cached content to the vehicle through one-hop downlink transmission.
	Thus, the service latency can be reduced compared with the conventional scheme.
	However, the contents cached at RSUs can go stale before next update.
	Assume the source can always generate contents to reflect the up-to-date status, i.e., zero AoI at the publishers.
	The AoI of a cached item will be set to the uplink transmission time once updated at the RSU, as shown in Fig.~\ref{fig_AoI}.
	Then, the AoI increases with time until updated again.
	The bandwidth is also split into $\beta B$ and $(1-\beta) B$ for the uplink cache update and content delivery, respectively.
	Intuitively, increasing $\beta$ can improve content freshness, while the delivery latency will increase.
	Therefore, the bandwidth splitting ratio $\beta$ is the key to tune the trade-off between AoI and service latency, which will be analyzed in the next section.
	
	\subsection{Request-Adaptive Scheme}
	
	The ReA scheme is introduced to further exploit the request arrival information.
	When a vehicle raises a request of item $s$, $s$ will be updated with probability $p_s$ at the RSU cache.
	If the update is not triggered, the RSU will directly send the cached content to the vehicle.
	Otherwise, the RSU fetches the new version from the publisher and then delivers the content.
	Under the ReA scheme, the update frequency of source $s$ depends on both update probability $p_s$ and request arrival rate $\lambda_s$, given by $p_s \lambda_s$.
	Compared with the RSUC scheme, the ReA scheme conducts finer-grained update control based on the content popularity.
	%	For example, the unpopular content items with low request arrival rate will not be updated frequently.

%%%%%%%%%%%%%%%%%%%%%%%%%%%%%%%%%%%%%%%%%%%%%%%%%%%%%%%%%%%%%%%%%%%%%%%%%%%%%%%%%%%%%%%%%%%%%%%%%%%%
\section{Freshness and Latency Analysis}
\label{sec_proposed_analysis}
	
	%\begin{figure}[!t]
	%	\centering
	%	\includegraphics[width=6in]{scenario}
	%	\caption{Air-ground integrated vehicular network overview.}
	%	\label{fig_scenario}
	%\end{figure}
	
	%\begin{figure*}[!t]
	%	\centering
	%	\subfloat[] {\includegraphics[width=2.5in]{evaluation_D}}
	%	\hfil
	%	\subfloat[]{\includegraphics[width=2.5in]{evaluation_AoI}}
	%	\caption{Analytical results evaluation, (a) service latency, and (b) AoI of vehicle received contents (request arrival rate set to 4 /ms).}
	%	\label{fig_evaluation}
	%\end{figure*}
	%
	
	%\begin{subequations}
	%	\label{equ_P1}
	%	\begin{align}
	%	\min\limits_{\rho_s, \theta_s}~~&~~\sum_{s=1}^{S} W_s \left( H_s \bar{A}_s + (1-H_s) \bar{D}_s \right) \\
	%	\mbox{(P1)}~~~~~s.t.~~&~~\sum_{s=1}^{S} \rho_s = 1, \\
	%	&~~ \rho_s > 0,  s=1,2,...,S,\\
	%	&~~ \theta_s \in \{2,3,4...\},  s=1,2,...,S,
	%	\end{align}
	%\end{subequations}
	This section analyzes the performances of three schemes.
	% regarding  the average AoI and service latency.
	
	\subsection{Conventional Scheme Analysis}
	
	% delay analysis
	The service process of the conventional scheme is shown in Fig.~\ref{fig_queue_conventional}. The the service time of both servers are considered to follow independent exponential distributions to reflect the randomness of wireless transmission.
	Accordingly, the queueing system of Fig.~\ref{fig_queue_conventional} is a Jackson queueing network, and the equilibrium joint probability distribution of queue lengths has a product-form solution.
	Thus, the uplink and downlink transmissions can be analyzed independently as two independent M/M/1 queues with equivalent arrival rate $\Lambda=\sum_{s=1}^{S} \lambda_s$. 
	Therefore, the average latency to serve a content request is given by
	\begin{equation}
	\label{eq_delay_con_1}
	\begin{split}
	\bar{D}_\mathrm{con} & = \frac{1}{\mu_\mathrm{UL}-\Lambda} + \frac{1}{\mu_\mathrm{DL}-\Lambda}\\
	& = \frac{1}{\beta R_\mathrm{UL}-\Lambda} +\frac{1}{(1-\beta) R_\mathrm{DL} -\Lambda},
	\end{split}
	\end{equation}
	where the two parts correspond to cache update and content delivery, respectively.
	The average latency $\bar{D}_\mathrm{con}$ increases with the total traffic arrival rate $\Lambda$ and decreases with the normalized uplink and downlink transmission rates.
	Furthermore, the bandwidth splitting ratio $\beta$ also influences the latency, and can be optimized by solving the following problem:
	\begin{subequations}
		\label{eq_P_band_opt_con}
		\begin{align}
		\mbox{(P1)}~~\min\limits_{\beta}~~~~&~~~~\frac{1}{\beta R_\mathrm{UL}-\Lambda} +\frac{1}{(1-\beta) R_\mathrm{DL} -\Lambda},\\
		s.t.~~~&~~~~\beta R_\mathrm{UL} \geq \Lambda, \\
		&~~~~(1-\beta) R_\mathrm{DL} \geq \Lambda, \\
		&~~~~0\leq\beta\leq 1,
		\end{align}
	\end{subequations}
	where the objective is to minimize the average latency, constraints (\ref{eq_P_band_opt_con}b) and (\ref{eq_P_band_opt_con}c) guarantee the stability of the uplink and downlink queues, respectively.
	By analyzing problem (P1), we obtain two important performance metrics: (1) the network capacity, i.e., the maximal request arrival rate that can be accommodated, and (2) the minimal average service latency with the optimal bandwidth splitting, given as Theorem~1.\\
	
	\textbf{Theorem 1.} Under the conventional scheme, the network capacity $\hat{\Lambda}_\mathrm{con}$ is given by
	\begin{equation}
	\label{eq_conv_cap}
	\hat{\Lambda}_\mathrm{con} = \frac{1}{\frac{1}{R_\mathrm{UL}} + \frac{1}{R_\mathrm{DL}}},
	\end{equation}
	and the minimal average service latency for the given traffic arrival rate $\Lambda$ is
	\begin{equation}
	\label{eq_conv_delay}
	\bar{D}_\mathrm{con}^* = \frac{\left(\frac{1}{\sqrt{R_\mathrm{UL}}}+ \frac{1}{\sqrt{R_\mathrm{DL}}}\right)^2}{1-\Lambda\left(\frac{1}{R_\mathrm{UL}} + \frac{1}{R_\mathrm{DL}}\right)}.
	\end{equation}
	\\

	\textit{Proof.}~Constraints (\ref{eq_P_band_opt_con}b) and (\ref{eq_P_band_opt_con}c) can be written as
	%		\begin{equation}
	%			\begin{split}
	%				& \frac{\Lambda}{R_\mathrm{UL}} \leq \beta, \\
	%				& \frac{\Lambda}{R_\mathrm{DL}} \leq 1-\beta.
	%			\end{split}
	%		\end{equation}
	\begin{equation}
	\begin{split}
	& \frac{\Lambda}{R_\mathrm{UL}} \leq \beta, \mbox{~~and~~} \frac{\Lambda}{R_\mathrm{DL}} \leq 1-\beta.
	\end{split}
	\end{equation}
	Therefore, 
	\begin{equation}
	\label{eq_P1_stability}
	\frac{\Lambda}{R_\mathrm{UL}} + \frac{\Lambda}{R_\mathrm{DL}} \leq 1,
	\end{equation}
	which is equivalent to
	\begin{equation}
	\Lambda \leq \frac{1}{\frac{1}{R_\mathrm{UL}} + \frac{1}{R_\mathrm{DL}}},
	\end{equation}
	revealing the system capacity.
	
	(P1) is a convex optimization problem with respect to $\beta$, which can be solved by applying the method of Lagrange multipliers.
	The Lagrange function is given by
	\begin{equation}
	\begin{split}
	& L\left(\beta, \nu_1, \nu_2, \nu_3, \nu_4\right) = \frac{1}{\beta R_\mathrm{UL}-\Lambda} +\frac{1}{(1-\beta) R_\mathrm{DL} -\Lambda} - \nu_1  \\
	&  \cdot (\beta R_\mathrm{UL}-\Lambda) - \nu_2 ((1-\beta) R_\mathrm{DL}-\Lambda) - \nu_3 \beta - \nu_4 (1-\beta),
	\end{split}
	\end{equation} 
	where $\nu_1$, $\nu_2$, $\nu_3$, and $\nu_4$ are the Lagrange multipliers, $\nu_1(\beta R_\mathrm{UL}-\Lambda) =0$, $\nu_2 ((1-\beta) R_\mathrm{UL}-\Lambda)=0$, $\nu_3 \beta=0$ and $\nu_4 (1-\beta)=0$.
	Take derivative with respect to $\beta$, we obtain the optimal condition of problem (P1):
	\begin{equation}
	\label{eq_P1_opt_condition}
	\begin{split}
	& -\frac{R_\mathrm{UL}}{(\beta R_\mathrm{UL} -\Lambda)^2} + \frac{R_\mathrm{DL}}{((1-\beta)R_\mathrm{DL} -\Lambda)^2} \\
	& - \nu_1 R_\mathrm{UL} + \nu_2 R_\mathrm{DL} - \nu_3 + \nu_4 = 0. 
	\end{split}
	\end{equation}
	Denote by 
	\begin{equation}
	\beta^*= \frac{\sqrt{R_\mathrm{UL} R_\mathrm{DL}} + \Lambda\left(1-\sqrt{\frac{R_\mathrm{UL}}{R_\mathrm{DL}}}\right)}{R_\mathrm{UL}+\sqrt{R_\mathrm{UL}R_\mathrm{DL}}},
	\end{equation}
	which satisfies (\ref{eq_P1_opt_condition}). 
	We can prove $\beta^*$ is feasible.
	\begin{equation}
	\begin{split}
	& \beta^* - \frac{\Lambda}{R_\mathrm{UL}} =  \frac{\sqrt{ R_\mathrm{DL}} + \frac{\Lambda}{\sqrt{R_\mathrm{UL}}} -\frac{\Lambda}{\sqrt{R_\mathrm{DL}}}}{\sqrt{R_\mathrm{UL}}+\sqrt{R_\mathrm{DL}}} -\frac{\Lambda}{R_\mathrm{UL}}\\
	& =   \frac{\sqrt{ R_\mathrm{DL}} \left[1 + \frac{\Lambda}{\sqrt{R_\mathrm{UL}R_\mathrm{DL}}} -\frac{\Lambda}{{R_\mathrm{DL}}} -\frac{\Lambda}{\sqrt{R_\mathrm{UL}R_\mathrm{DL}}} - \frac{\Lambda}{R_\mathrm{UL}}\right]}{\sqrt{R_\mathrm{UL}}+\sqrt{R_\mathrm{DL}}}\\
	& = \frac{\sqrt{ R_\mathrm{DL}} \left[1 -\frac{\Lambda}{{R_\mathrm{DL}}} - \frac{\Lambda}{R_\mathrm{UL}}\right]}{\sqrt{R_\mathrm{UL}}+\sqrt{R_\mathrm{DL}}} \geq 0,
	\end{split}
	\end{equation}
	according to (\ref{eq_P1_stability}). Thus, constraint (\ref{eq_P_band_opt_con}b) holds.
	In addition,
	\begin{equation}
	\begin{split}
	& \beta^* -1 +\frac{\Lambda}{R_\mathrm{DL}}  \\
	= & \frac{-\sqrt{R_\mathrm{UL}} + \frac{\Lambda}{\sqrt{R_\mathrm{UL}}}-\frac{\Lambda}{\sqrt{R_\mathrm{DL}}} +\frac{\sqrt{R_\mathrm{UL}}}{R_\mathrm{DL}}\Lambda + \frac{\Lambda}{\sqrt{R_\mathrm{DL}} }}{\sqrt{R_\mathrm{UL}}+\sqrt{R_\mathrm{DL}}} \\
	= & \frac{\sqrt{R_\mathrm{UL}} \left[ \frac{\Lambda}{{R_\mathrm{UL}}} + \frac{\Lambda}{R_\mathrm{DL}} -1\right] }{\sqrt{R_\mathrm{UL}}+\sqrt{R_\mathrm{DL}}} \leq 0, \\
	\end{split}
	\end{equation}		
	and constraint (\ref{eq_P_band_opt_con}c) holds.
	Therefore, $\beta^*$ is the optimal solution to (P1). 
	Substituting $\beta^*$ into (\ref{eq_delay_con_1}), the minimal average latency can be obtained, given as Theorem 1.  			
	\hfill \rule{4pt}{8pt}\\

	% AoI: analysis based on the optimal bandwidth allocation	
	According to the conventional scheme, the AoI of vehicle-received content includes two parts: (1) the time of uplink transmission, and (2) the dwelling time at the downlink channel including queueing and transmission\footnote{The uplink queuing time does not influence the AoI, since the RSU fetches the new content version when starting to serve the user.}.
	Thus, the average AoI at vehicle side is given by
	\begin{equation}
	\footnotesize
	\label{eq_con_AoI}
	\begin{split}
	& \bar{A}_\mathrm{con}  =  \frac{1}{\beta^* R_\mathrm{UL}} + \frac{1}{R_\mathrm{DL}(1-\beta^*)-\Lambda}\\
	& \!=\!  \frac{1\!+\!\sqrt{\frac{R_\mathrm{DL}}{R_\mathrm{UL}}}}{\sqrt{R_\mathrm{UL}R_\mathrm{DL}} \!+\!\Lambda\left(1\!-\!\sqrt{\frac{R_\mathrm{UL}}{R_\mathrm{DL}}}\right)} \!+\! \frac{1\!+\!\sqrt{\frac{R_\mathrm{UL}}{R_\mathrm{DL}}}}{\sqrt{R_\mathrm{UL}R_\mathrm{DL}}\!-\!\Lambda\left(\sqrt{\frac{R_\mathrm{UL}}{R_\mathrm{DL}}}\!+\!\sqrt{\frac{R_\mathrm{DL}}{R_\mathrm{UL}}}\right)}.
	\end{split}
	\end{equation} 
	The average AoI increases with traffic arrival rate, which can be proved by taking derivative of $\bar{A}_\mathrm{con}$ with respect to $\Lambda$.
	This is reasonable as the queue length of the downlink channel increases with traffic load, which is also responsible for the staleness of information.	
	
	\textbf{Corollary 1.} According to the derived capacity, service latency and AoI, the conventional scheme has following features regarding system performance.
	%		\begin{itemize}
	%			\item The capacity depends on the normalized service rates of both uplink and downlink channels;\\
	%			\item The average AoI and latency do not have trade-off relationship, and both increase with traffic load;\\
	%			\item The channel unsymmetry degrades the network capacity and service latency.
	%		\end{itemize}
	(1) The capacity depends on the normalized service rates of both uplink and downlink channels;
	(2) The average AoI and latency both increase with traffic load;
	and (3) The channel unsymmetry degrades the network capacity and service latency.
	
	\subsection{RSU-Centric Scheme Analysis}
	
	% Delay analysis
	Under the RSUC scheme, vehicles obtain the requested content through one-hop downlink transmission.
	The average service latency is thus given by
	\begin{equation}
	\label{eq_RC_delay}
	\bar{D}_\mathrm{RC} = \frac{1}{(1-\beta)R_\mathrm{DL}-\Lambda},
	\end{equation}
	which increases with traffic load and bandwidth splitting ratio.

	% AoI analysis
	The analysis of average AoI is more challenging. 
	From the RSU side, the content update of a typical item $s$ can be modeled as a renewal process.
	The epoch happens when item $s$ is updated in the RSU cache.
	Denote by $T_{\mathrm{update}}$ the update interval, i.e., the time duration between two successive content updates of the same item, given by
	\begin{equation}
	T_{\mathrm{update}} = \sum_{s=1}^{S} t_{s},
	\end{equation}
	where $t_s$ is the uplink transmission time when updating item $s$.
	$t_1,t_2,\cdots,t_S$ are independent and identically distributed random variables following exponential distribution of ${1}/{\beta R_\mathrm{UL}}$.
	Thus, the update cycle $T_\mathrm{update}$ follows Erlang-$S$ distribution, and
	\begin{equation}
	\begin{split}
	& \mathds{E}\left[T_\mathrm{update} \right] = \frac{S}{\beta R_\mathrm{UL}},\\
	& \mathds{E} \left[T_\mathrm{update}^2 \right] -  \mathds{E}^2 \left[T_\mathrm{update} \right] = \frac{S}{(\beta R_\mathrm{UL})^2}.
	\end{split}
	\end{equation}
	Suppose a vehicle requests item $s$ and the downlink transmission begins at time $t$.
	The AoI of the received content is the summation of two parts: (1) the AoI of item $s$ in the RSU cache at time $t$, and (2) the downlink transmission time for content delivery.
	For the RSU cache, the AoI of item $s$ at time $t$ also includes two parts: (1) the uplink transmission time to update source $s$ at the last epoch, and (2) the spent time, i.e., the time between $t$ and the last epoch.
	Denote by $T_\mathrm{b}$ the spent time, given by 
	\begin{equation}
	\begin{split}
	\mathds{E}[T_\mathrm{b}] & = \frac{\mathds{E} \left[T_\mathrm{update}^2 \right]}{2 \mathds{E}\left[T_\mathrm{update} \right]} = \frac{1+S}{2\beta R_\mathrm{UL}},
	\end{split}
	\end{equation}
	according to the theories of random process.
	Denote by $T_\mathrm{UL}$ and $T_\mathrm{DL}$ the uplink and downlink transmission time, respectively.
	Thus, the average AoI of vehicle received content can be obtained:
	\begin{equation}
	\label{eq_RC_AoI}	
	\begin{split}
	& \bar{A}_\mathrm{RC}  = \mathds{E}[T_\mathrm{UL}] + \mathds{E}[T_\mathrm{b}] + \mathds{E}[T_\mathrm{DL}],\\
	& = \frac{1}{\beta R_\mathrm{UL}} + \frac{1+S}{2\beta R_\mathrm{UL}} + \frac{1}{(1-\beta) R_\mathrm{DL}} \\
	& = \frac{S+3}{2\beta R_\mathrm{UL}} + \frac{1}{(1-\beta) R_\mathrm{DL}}
	\end{split}
	\end{equation}
	which is a convex function with respect to $\beta$.\\

	\textbf{Theorem~2.} Under the RSUC scheme, the AoI and service latency have a trade-off if the ratio of bandwidth allocated to the uplink is no larger than a certain threshold:
	\begin{equation}
	\beta \leq \frac{1}{\sqrt{\frac{2R_\mathrm{UL}}{(S+3)R_\mathrm{DL}}}+1}.
	\end{equation}
	Otherwise, the AoI and service latency both increase with $\beta$.
	
	\textit{Proof.} By taking the first- and second-order derivatives of Eq.~(\ref{eq_RC_AoI}) with respect to $\beta$, we can prove that the average AoI is a convex function of $\beta$, with global minimum of $1/\left(\sqrt{\frac{2R_\mathrm{UL}}{(S+3)R_\mathrm{DL}}}+1\right)$. According to Eq.~(\ref{eq_RC_delay}), the service latency can be proved to increase with $\beta$ in a monotone manner. Theorem~2 is thus proved. \hfill \rule{4pt}{8pt}\\
	
	Different from the intuition, Theorem 2 indicates that content freshness and service latency do not always have a trade-off. 
	In particular, frequently updating the cached can even degrade the freshness of vehicle-received content, in addition to increasing the service latency.
	The reason is as follows.
	When the bandwidth allocated to uplink is small (i.e., small $\beta$), increasing $\beta$ can significantly improve the freshness of cached contents at the RSU, at the cost of reducing the downlink transmission rate.
	Thus, the AoI and service latency show a trade-off relationship with respect to bandwidth splitting ratio $\beta$.
	When sufficient bandwidth is allocated for content update (i.e., large $\beta$), further increasing $\beta$ will no longer improve the freshness of cached contents due to marginal gain effect, whereas the extremely low transmission rate in the downlink will stale the contents delivered to vehicles.\\	
	
	\textbf{Corollary 2.} Under the RSU-centric scheme, the bandwidth splitting ratio should be adjusted to balance the AoI and service latency, in the range of $\left[0,\frac{1}{\sqrt{\frac{2R_\mathrm{UL}}{(S+3)R_\mathrm{DL}}}+1}\right)$. \\
	
	The threshold of $\beta$ can be interpreted as saturate point, which indicates the bandwidth allocated to the uplink is sufficient to maintain content freshness.
	For networks with low uplink transmission rates or large number of sources, the threshold is high and more bandwidth can be allocated to the uplink for content update.

	% Delay and AoI tradeoff analysis
	
	\subsection{Request-Adaptive Scheme Analysis}
	
	% Delay analysis	
	The service process of request-adaptive scheme can be modeled by a M/G/1 queue with the arrival rate of $\Lambda$.
	Denote by $X_1$ and $X_2$ the uplink and downlink transmission time following exponential distributions of $1/R_\mathrm{UL}$ and $1/R_\mathrm{DL}$,  respectively\footnote{Under the request-adaptive scheme, the uplink and downlink share the spectrum resource in the time domain without bandwidth splitting.}.
	Denote by $I$ a 0-1 indicator showing if the content update is triggered.
	Thus, the service time $X=X_2$ if $I=0$, while $X=X_1+X_2$ otherwise.
	As $X_1$ and $X_2$ are independent, 
	\begin{equation}
	\begin{split}
	\mathds{E}[X] &= P \left( \mathds{E}[X_1]\!+\!\mathds{E}[X_2] \! \right) \!+\! (\!1\!-\!P\!)\! \left(\mathds{E}[X_2]\right) \!=\! \frac{P}{R_\mathrm{UL}} \!+\! \frac{1}{R_\mathrm{DL}}\\
	\mathds{E}[X^2] &= P \left( \mathds{E}[(X_1+X_2)^2] \right) + (1-P) \left(\mathds{E}[X_2^2]\right) \\
	& = P \mathds{E}[X_1^2] + 2 P \mathds{E}[X_1]\mathds{E}[X_2]  + \mathds{E}[X_2^2]\\
	& = \frac{2P}{R_\mathrm{UL}^2} + \frac{2}{R_\mathrm{DL}^2} +\frac{2P}{R_\mathrm{UL}R_\mathrm{DL}},
	\end{split}
	\end{equation}
	where $P = \frac{\sum_{s=1}^{S} p_s \lambda_s}{\sum_{s=1}^{S} \lambda_s}$.
	$P$ is defined as update ratio, denoting the average probability that a content request triggers cache update.
	%	Therefore, the variance of service time is given by
	%		\begin{equation}
	%			var[X] = \mathds{E}[X^2] - \mathds{E}^2[X] = \frac{1}{R_\mathrm{UL}^2} +\frac{1}{R_\mathrm{DL}^2} -\frac{(1-P)^2}{R_\mathrm{UL}^2}.
	%		\end{equation}
	Applying the queueing theory, the average service latency of ReA scheme can be obtained:
	\begin{equation}
	\label{eq_RA_delay}
	\begin{split}
	\bar{D}_\mathrm{RA} & = \frac{\Lambda \mathds{E}[X^2] }{2(1-\Lambda \mathds{E}[X])} + \mathds{E}[X] \\%= \frac{\rho h \left(1+C_b^2 \right)}{2(1-\rho)} + h 
	& = \frac{\Lambda\left[ \frac{P}{R_\mathrm{UL}^2} + \frac{1}{R_\mathrm{DL}^2} +\frac{P}{R_\mathrm{UL}R_\mathrm{DL}} \right]}{1-\Lambda\left(\frac{P}{R_\mathrm{UL}}+\frac{1}{R_\mathrm{DL}}\right)} + \frac{P}{R_\mathrm{UL}} +\frac{1}{R_\mathrm{DL}} \\
	& = \frac{\frac{1}{R_\mathrm{DL}}+\frac{P\Lambda}{R_\mathrm{UL}^2}}{1-\Lambda \left( \frac{P}{R_\mathrm{UL}} + \frac{1}{R_\mathrm{DL}} \right)} +\frac{P}{R_\mathrm{UL}},
	\end{split}
	\end{equation}	
	which increases with $P$ (i.e., frequent cache update) and $\Lambda$ (i.e., heavy traffic load) and decreases with $R_\mathrm{UL}$ and $R_\mathrm{DL}$ (i.e., larger service rate).
	
	The network capacity can also be obtained from (\ref{eq_RA_delay}):
	\begin{equation}
	\hat{\Lambda}_\mathrm{RA} = \frac{1}{\frac{P}{R_\mathrm{UL}}+\frac{1}{R_\mathrm{DL}}}. %注意：等式右边的P和容量也有关系，不算是闭式容量表达式
	\end{equation}
	When $P=1$, the ReA scheme has the equivalent capacity but lower latency compared with the conventional scheme.
	In addition, the ReA scheme can always outperform the conventional scheme regarding capacity and latency, since $\hat{\Lambda}_\mathrm{RA}$ decreases with the update ratio $P$.
	Specifically, when $P=0$, the RSU cache never updates and all bandwidth is used for content delivery (i.e., static caching).
	The capacity increases to the normalized downlink transmission rate $R_\mathrm{DL}$, and the service latency decreases to $\frac{1}{R_\mathrm{DL}-\Lambda}$.
	When $P$ varies within range $(0,1)$, the network capacity and service latency can be traded by sacrificing content freshness.
	
	% AoI analysis:可能有问题
	The AoI performance can be analyzed in the similar way as the RSUC scheme, where the update process of cached item $s$ can also be modeled as a renewal process.
	However, the update interval is completely different:
	\begin{equation}
	T_{\mathrm{update},s} = \sum_{k=1}^{N_s} T_{k},
	\end{equation}
	where $N_s$ denotes the number of requests served between two successive updates of item $s$, and $T_k$ is the time duration between the departures of the $k$th and $(k-1)$th requests\footnote{The 0th request departs at time 0.}.
	$T_k$ are i.i.d. random variables following exponential distribution with mean $1/\lambda_s$, applying the properties of M/G/1 queue.
	Under the ReA scheme, $N_s$ follows geometric distribution: $\mathds{P}[N_s=n]=p_s(1-p_s)^{n-1}$.
	Therefore, 
	\begin{equation}
	\mathds{E}[T_\mathrm{update,s}] = \sum_{n=1}^{\infty} p_s(1-p_s)^{n-1} \frac{n}{\lambda_s} = \frac{1}{p_s \lambda_s},
	\end{equation}
	\begin{equation}
	\mathds{E}[T_\mathrm{updat,s}^2] = \sum_{n=1}^{\infty} p_s(1-p_s)^{n-1} \frac{n^2+n}{\lambda_s^2} = \frac{2}{\lambda_s^2 p_s^2},
	\end{equation}
	and the spent time is given by
	\begin{equation}
	\mathds{E}[T_\mathrm{b}] = \frac{\mathds{E} \left[T_\mathrm{update}^2 \right]}{2 \mathds{E}\left[T_\mathrm{update} \right]} = \frac{1}{\lambda_s p_s}.
	\end{equation}
	If the request for item $s$ triggers update, the average AoI of vehicle received content is given by $\frac{1}{R_\mathrm{UL}}+\frac{1}{R_\mathrm{DL}}$.
	Otherwise, the average AoI equals to $\mathds{E}{[T_\mathrm{b}}] +\frac{1}{R_\mathrm{DL}}$.
	As the request triggers update with probability $p_s$, the overall average AoI of vehicle received content is given by
	\begin{equation}
	\label{eq_RA_AoI}
	\begin{split}
	\bar{A}_{\mathrm{RA},s} \!&\! =\! p_sc\! \left(\!\frac{1}{R_\mathrm{UL}}\!+\!\frac{1}{R_\mathrm{DL}}\!\right)\! +\! (\!1\!-\!p_s\!)\! \left(\!\mathds{E}[T_\mathrm{b}] +\frac{1}{R_\mathrm{DL}}\!\right)\\
	& = \frac{p_s}{R_\mathrm{UL}} +\frac{1}{R_\mathrm{DL}} + \frac{1-p_\mathrm{s}}{p_s\lambda_s}.
	\end{split}
	\end{equation}
	Based on the derived analytical results, the trade-off between AoI and service latency can be obtained.\\
	
	\textbf{Theorem 3.} Under the ReA scheme, the AoI and service latency always have a trade-off if $R_\mathrm{UL}\geq \lambda_s$ for $s=1,2,...,S$.
	Otherwise, the trade-off only holds in the region of $p_s\leq \sqrt{R_\mathrm{UL}/\lambda_s}$, $\forall s$.
	
	\textit{Proof.} The average AoI is a convex function of $p_s$ with a global minimum of $\sqrt{R_\mathrm{UL}/\lambda_s}$. If $R_\mathrm{UL} \geq \lambda_s$, the average AoI decreases with $p_s\in[0,1]$. As the average latency increases with $p_s$, the average AoI and latency have a trade-off. \hfill \rule{4pt}{8pt}\\
	
	Notice that the content freshness and service latency cannot be traded off if the uplink is overloaded, i.e., $R_\mathrm{UL} < \lambda_s$.
	This may happen due to non-ideal uplink transmissions, e.g., low-power sensor nodes.
	In this case, restraining the update frequency can enhance both service latency and content freshness, and the update probability should not exceed $\sqrt{R_\mathrm{UL}/\lambda_s}$.
	On the contrary, the freshness and latency can always be traded off if $R_\mathrm{UL} \geq \lambda_s$.
	Here are two extreme cases.
	
	\textbf{Case 1.} As $p_s\rightarrow0$ (i.e., the static caching without update), $\bar{A}_{\mathrm{RA},s} \rightarrow \infty$. All resources are utilized for content delivery, which achieves the minimal service latency $\frac{1}{R_\mathrm{DL}-\Lambda}$.
	
	\textbf{Case 2.} When $p_s=1$, all requests trigger update. The average service latency achieves the maximum, given by
	\begin{equation}
	\frac{\frac{1}{R_\mathrm{DL}}+\frac{1}{R_\mathrm{UL}}-\frac{\Lambda}{R_\mathrm{DL}R_\mathrm{UL}}} {1-\Lambda\left(\frac{1}{R_\mathrm{UL}} +\frac{1}{R_\mathrm{DL}} \right)}.
	\end{equation}
	The corresponding average AoI is given by $1/{R_\mathrm{UL}} +1/{R_\mathrm{DL}}$, which is the limit of content freshness that can be achieved.

%%%%%%%%%%%%%%%%%%%%%%%%%%%%%%%%%%%%%%%%%%%%%%%%%%%%%%%%%%%%%%%%%%%%%%%%%%%%%%%%%%%%%%%%%%%%%%%%%%%%
\section{RSUC and ReA Scheme Optimization}
\label{sec_tradeoff}
	In this section, the RSUC and ReA schemes are further optimized to enhance the freshness and latency performances simultaneously.
	
	\subsection{RSUC Scheme Optimization}
	
	Based on Eqs.~(\ref{eq_RC_delay}) and (\ref{eq_RC_AoI}), the RSUC optimization problem can be formulated:
	\begin{subequations}
		\label{eq_P_RC_weighted}
		\begin{align}
		\mbox{(P2)}~\min\limits_{\beta}~&~\frac{1}{\!(\!1\!-\!\beta\!)\! R_\mathrm{DL}\!-\!\Lambda} \!+\! W_\mathrm{A} \!\left[\! \frac{S\!+\!3}{2\beta R_\mathrm{UL}} \!+\! \frac{1}{\!(\!1\!-\!\beta)\!R_\mathrm{DL}}\! \right]\!,\\
		s.t.&~(1-\beta) R_\mathrm{DL} \geq \Lambda, \\
		&~0\leq\beta\leq 1,
		\end{align}
	\end{subequations}
	where $W_\mathrm{A}$ is a weight factor, indicating the importance of AoI compared with service latency.
	The objective is to minimize both the latency and AoI.
	The first constraint (\ref{eq_P_RC_weighted}) guarantees that the downlink is not overloaded.
	Problem (P2) is a convex optimization problem, which can be addressed by the method of Lagrange multipliers.
	The optimal condition is given by:
	\begin{equation}
	\label{eq_RC_weighted_opt_condition}
	\frac{1}{\left(\frac{1-\frac{\Lambda}{R_\mathrm{DL}}}{\beta^*}-1\right)^2} +\frac{W_\mathrm{A}}{\left(\frac{1}{\beta^*}-1\right)^2} = \frac{(S+3)W_\mathrm{A} R_\mathrm{DL}}{2 R_\mathrm{UL}}.
	\end{equation}
	%while the derivation details are omitted here due to page limit.
	Note that $0\leq \beta^* \leq 1-\frac{\Lambda}{R_\mathrm{DL}}$, based on the constraint (\ref{eq_P_RC_weighted}b).
	Accordingly, the left part of (\ref{eq_RC_weighted_opt_condition}) increases with $\beta^*$, varying in range of $[0,\infty)$.
	Therefore, (\ref{eq_RC_weighted_opt_condition}) has a unique solution, which can be obtained by dichotomy searching.
	
	The optimal bandwidth splitting ratio $\beta^*$ achieves the Pareto-optimality of AoI and latency, which further depends on system parameters.
	In specific, $\beta^*$ increases with $S$ but decreases with $R_\mathrm{UL}$.
	This result indicates that more resources are consumed to maintain the freshness of more content items, or in case of bad uplink channel conditions.
	Furthermore, $\beta^*$ increases with the weight factor $W_\mathrm{A}$, revealing the optimal AoI-latency trade-off.
	
	Problem (P2) mainly applies to elastic AoI requirements.
	For the applications with strict average AoI requirements, the problem can be revised as follows:
	\begin{subequations}
		\label{eq_P_RC_strict}
		\begin{align}
		\mbox{(P3)}~\min\limits_{\beta}~&~~\frac{1}{(1-\beta) R_\mathrm{DL}-\Lambda},\\
		s.t.&~~(1-\beta) R_\mathrm{DL} \geq \Lambda, \\
		&~~~\frac{S+3}{2\beta R_\mathrm{UL}} + \frac{1}{(1-\beta)R_\mathrm{DL}} \leq \hat{A}_\mathrm{RC},\\
		&~~~0\leq\beta\leq 1,
		\end{align}
	\end{subequations}
	where $\hat{A}_\mathrm{RC}$ is the threshold for average AoI.\\
	
	\textbf{Theorem 4.} If the traffic load satisfies $\Lambda < \frac{R_\mathrm{DL}}{1+\sqrt{\frac{(S+3)R_\mathrm{DL}}{2R_\mathrm{UL}}}}$, the average AoI achieved by RSUC scheme satisfies
	\begin{equation}
	\label{eq_AoI_min}
	\hat{A}_\mathrm{RC} \geq \left( \frac{1}{\sqrt{R_\mathrm{DL}}} + \sqrt{\frac{S+3}{2 R_\mathrm{UL}}}  \right) ^2.
	\end{equation}
	The equality holds if the bandwidth splitting ratio is set to 
	\begin{equation}
	\label{eq_theorem_4}
	\hat{\beta}=1-\frac{1}{1+\sqrt{\frac{(S+3) R_\mathrm{DL}}{2R_\mathrm{UL}}}}.
	\end{equation}
	
	\textit{Proof.} According to the constraint (\ref{eq_P_RC_strict}c), we have
	\begin{equation}
	\label{eq_AoI_min_analysis_0}
	\frac{S+3}{2 \beta R_\mathrm{UL}} \leq \hat{A}_\mathrm{RC} \beta - \frac{\beta}{(1-\beta) R_\mathrm{DL}}
	\end{equation}
	where the right part can be rewritten as
	\begin{subequations}
		\label{eq_AoI_min_analysis}
		\begin{align}
		& -\hat{A}_\mathrm{RC} (1-\beta)-\frac{1}{(1-\beta)R_\mathrm{DL}}+ \hat{A}_\mathrm{RC} + \frac{1}{R_\mathrm{DL}},\\
		\leq & -2\sqrt{\frac{\hat{A}_\mathrm{RC}}{R_\mathrm{DL}}} +\hat{A}_\mathrm{RC} +\frac{1}{R_\mathrm{DL}} \\
		= & \left( \sqrt{\hat{A}_\mathrm{RC}} -\frac{1}{\sqrt{R_\mathrm{DL}}} \right)^2,
		\end{align}
	\end{subequations}
	and the equality of (\ref{eq_AoI_min_analysis}b) holds if $1-\beta = \sqrt{\frac{1}{\hat{A}_\mathrm{RC} R_\mathrm{DL}}}$.
	According to the form of AoI, $\hat{A}_\mathrm{RC}>\frac{1}{R_\mathrm{DL}}$.
	Combining (\ref{eq_AoI_min_analysis_0}) and (\ref{eq_AoI_min_analysis}), we have
	\begin{equation}
	\sqrt{\hat{A}_\mathrm{RC}} - \frac{1}{\sqrt{R_\mathrm{DL}}} - \sqrt{\frac{S+3}{2R_\mathrm{UL}}} \geq 0,
	\end{equation}
	proving (\ref{eq_AoI_min}) in Theorem 3.
	Take the equality of (\ref{eq_AoI_min_analysis}), the equality of (\ref{eq_AoI_min}) holds.
	Substitute (\ref{eq_AoI_min}) into $1-\beta = \sqrt{\frac{1}{\hat{A}_\mathrm{RC} R_\mathrm{DL}}}$, and we obtain (\ref{eq_theorem_4}).
	Note that $0<\hat{\beta}<1$, and $(1-\hat{\beta})R_\mathrm{DL}\geq \Lambda$ under the condition of Theorem 4.
	Thus, $\hat{\beta}$ is feasible to (P3) and Theorem 4 is proved. \hfill \rule{4pt}{8pt}\\
	
	Theorem 4 reveals the limit of content freshness that can be achieved by the RSUC scheme. In specific, the content freshness degrades with the number of content items. 
	Furthermore, the content freshness is more sensitive to the uplink channel condition, especially with large number of content items.
	In addition, both the vehicular network capacity and service latency have trade-offs with the AoI constraint, as the minimal value of $\beta$ satisfying constraint (\ref{eq_P_RC_strict}c) decreases with $\hat{A}_\mathrm{RC}$.
	Here are two cases to show capacity-freshness trade-off.
	Under the most strict requirement of freshness (i.e., equality of (\ref{eq_AoI_min})), the network capacity is
	\begin{equation}
	\hat{\Lambda}_\mathrm{RC} = \frac{R_\mathrm{DL}}{1+\sqrt{\frac{(S+3)R_\mathrm{DL}}{2R_\mathrm{UL}}}},
	\end{equation} 
	with (\ref{eq_theorem_4}) in Theorem 4.
	On the contrary, without content freshness requirement  (i.e., $\hat{A}_\mathrm{RC} \rightarrow \infty$), the vehicular network capacity is given by $\hat{\Lambda}_\mathrm{RC} \rightarrow R_\mathrm{DL}$ with $\hat{\beta}=0$.
	%Thus, the network capacity takes in the range of $\left(\frac{R_\mathrm{DL}}{1+\sqrt{\frac{(S+3)R_\mathrm{DL}}{2R_\mathrm{UL}}}}, R_\mathrm{DL}\right)$.
	Thus, we can use the RSUC scheme to effectively improve vehicular network capacity in a wider range by sacrificing the content freshness, if there are large number of source nodes or the uplink channel suffers from deep fading.
	The AoI-latency trade-off can be analyzed in the similar manner.

	\subsection{AoI and Latency Trade-off under ReA Scheme}
	
	The ReA optimization problem can be formulated as follows\footnote{The optimization under elastic AoI requirements is omitted due to page limit. (P4) can provide more insights into system performance.}:
	\begin{subequations}
		\label{eq_P_RA_aver}
		\begin{align}
		\mbox{(P4)}~~\min_{p_s}~&~~\frac{\sum_{s=1}^{S} p_s \lambda_s}{\sum_{s=1}^{S} \lambda_s} \\
		s.t.~&~ \frac{\sum_{s=1}^{S} \lambda_s \bar{A}_{\mathrm{RA},s}}{\sum_{s=1}^{S} \lambda_s} \leq \hat{A}_\mathrm{RA},\\
		&~0\leq p_s\leq 1,~~s=1,2,...,S,
		\end{align}
	\end{subequations}
	where the objective is to minimize the overall content update ratio, constraint (\ref{eq_P_RA_aver}b) guarantees that the system-level average AoI not to exceed the threshold $\hat{A}_\mathrm{RA}$.
	As the average service latency increases with the content update ratio, (P4) minimizes the service latency under the given AoI requirement. 
	
	Substitute (\ref{eq_RA_AoI}) into (\ref{eq_P_RA_aver}b), we have
	\begin{equation}
	\label{eq_RA_ana_temp_0}
	\sum_{s=1}^{S}\left[\frac{ \lambda_s p_s}{R_\mathrm{UL}} +\frac{1}{p_s}\right] \leq S + \Lambda \left( \hat{A}_\mathrm{RA} -\frac{1}{R_\mathrm{DL}}\right).
	\end{equation}
	As (P4) is a convex optimization problem, we can apply Lagrange method to find the solution.
	The optimality condition is given by
	\begin{equation}
	\frac{\lambda_s}{\Lambda} + \frac{\nu_0 \lambda_s}{R_\mathrm{UL}} -\frac{\nu_0}{p_s^2} -\nu_1+\nu_2 =0,
	\end{equation}
	where $\nu_0$, $\nu_1$, and $\nu_2$ are the Lagrange multipliers. 
	Thus, 
	\begin{equation}
	\label{eq_RA_ana_temp_1}
	p_s = \frac{1}{\sqrt{ \frac{\lambda_s}{\Lambda \nu_0} + \frac{\lambda_s}{R_\mathrm{UL}} - \frac{\nu_1}{\nu_0} + \frac{\nu_2}{\nu_0}}}.
	\end{equation}
	Suppose $0<p_s<1$ and $\nu_1=\nu_2=0$. 
	Substituting (\ref{eq_RA_ana_temp_1}) into (\ref{eq_RA_ana_temp_0}) and taking equality, we can obtain
	\begin{equation}
	\label{eq_RA_ana_temp_2}
	2 \sqrt{\frac{1}{\nu_0 \Lambda} +\frac{1}{R_\mathrm{UL}}} = \left[Y \pm \sqrt{Y^2-\frac{4}{R_\mathrm{UL}}}\right],
	\end{equation}
	where 
	\begin{equation}
	Y = \frac{S+\Lambda \left(\hat{A}_\mathrm{RA} - \frac{1}{R_\mathrm{DL}} \right)}{\sum_{s=1}^{S} \sqrt{\lambda_s}},
	\end{equation}
	for notation simplicity.
	Note that $p_s$ should take the smaller value which satisfying the equality of (\ref{eq_RA_ana_temp_0}).
	Thus, $\pm$ should set to $+$ in (\ref{eq_RA_ana_temp_2}), and
	\begin{equation}
	\label{eq_P4_opt}
	\hat{p}_s = \frac{2}{\sqrt{\lambda_s} \left( Y + \sqrt{Y^2-\frac{4}{R_\mathrm{UL}}}\right)}.
	\end{equation}
	Accordingly, an iterative algorithm can be designed to solve (P4).
	Denote by $\mathcal{S}_\mathrm{sub}=\{s|\hat{p}_s \geq 1\}$.
	If $\mathcal{S}_\mathrm{sub}=\emptyset$, the optimal solution $p^*_s=\hat{p}_s$ for $s\in\mathcal{S}$, and (P4) is solved.
	Otherwise, $p^*_s=1$ for $s\in\mathcal{S}_\mathrm{sub}$.
	Set $\mathcal{S}=\mathcal{S}-\mathcal{S}_\mathrm{sub}$, substitute it into (P4), and calculate (\ref{eq_P4_opt}) in the new iteration.

%%%%%%%%%%%%%%%%%%%%%%%%%%%%%%%%%%%%%%%%%%%%%%%%%%%%%%%%%%%%%%%%%%%%%%%%%%%%%%%%%%%%%%%%%%%%%%%%%%%
\section{Simulation and Numerical Results}
\label{sec_simulation}
	
	\begin{figure}[!t]
		\centering
		\subfloat[] {\includegraphics[width=1.6in]{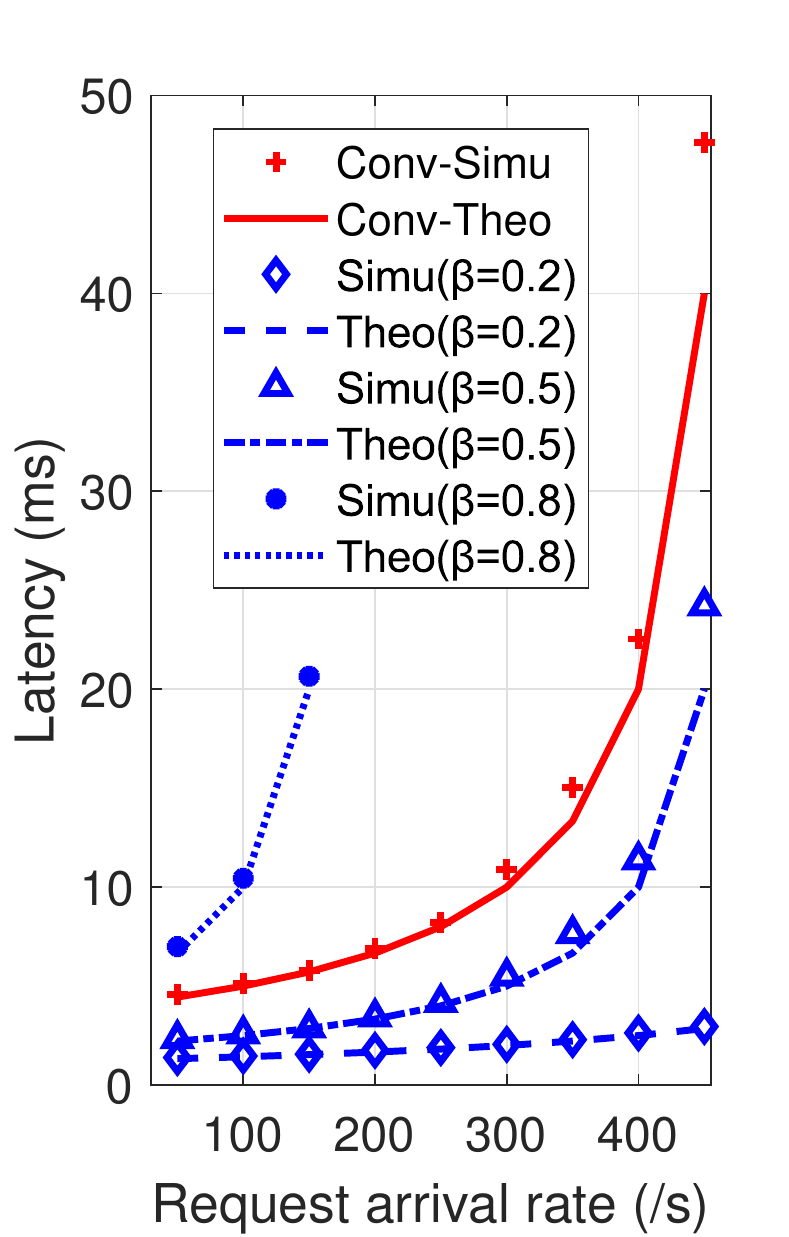}}
		%\hfil
		\subfloat[]{\includegraphics[width=1.6in]{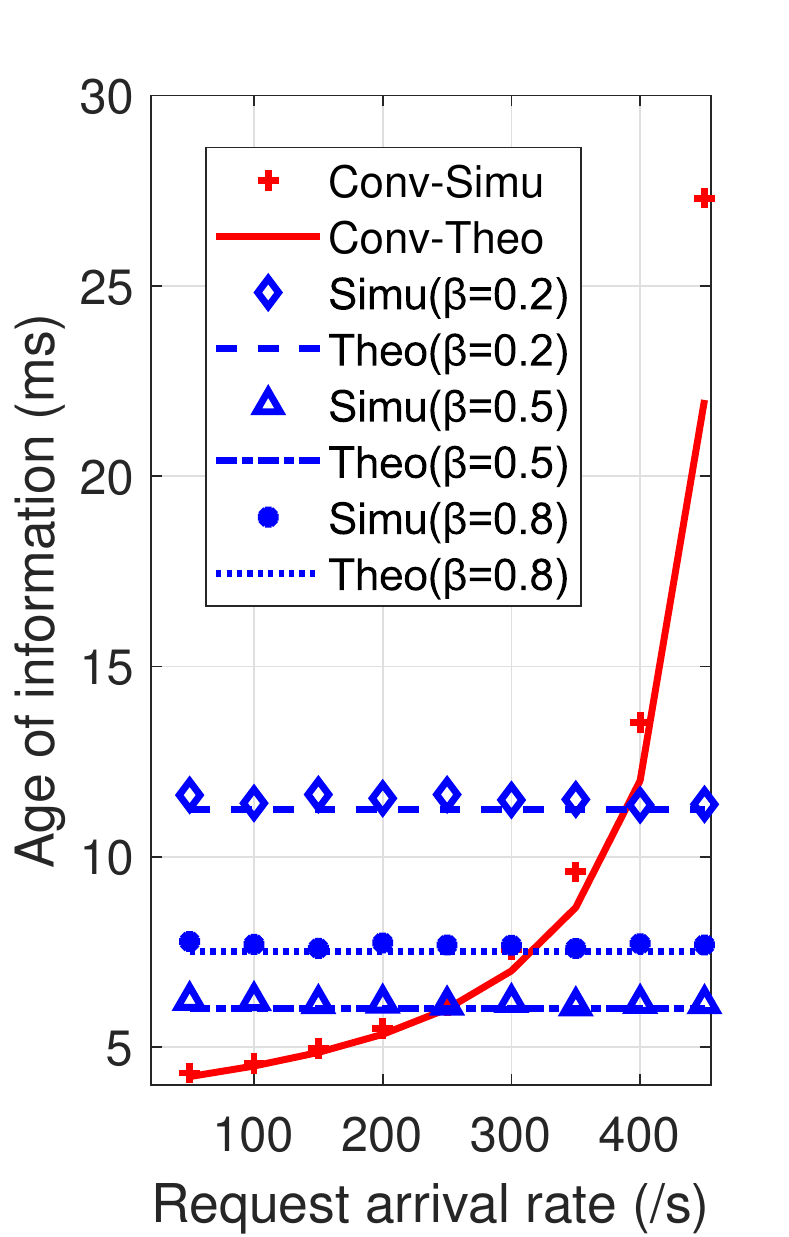}}
		%\hfil
		\caption{Analytical results validation of the RSUC scheme, with conventional scheme as baseline (a) service latency, (b) age of information.}
		\label{fig_RSUC_validation}
	\end{figure}
	
	\begin{figure}[!t]
		\centering
		\subfloat[] {\includegraphics[width=1.6in]{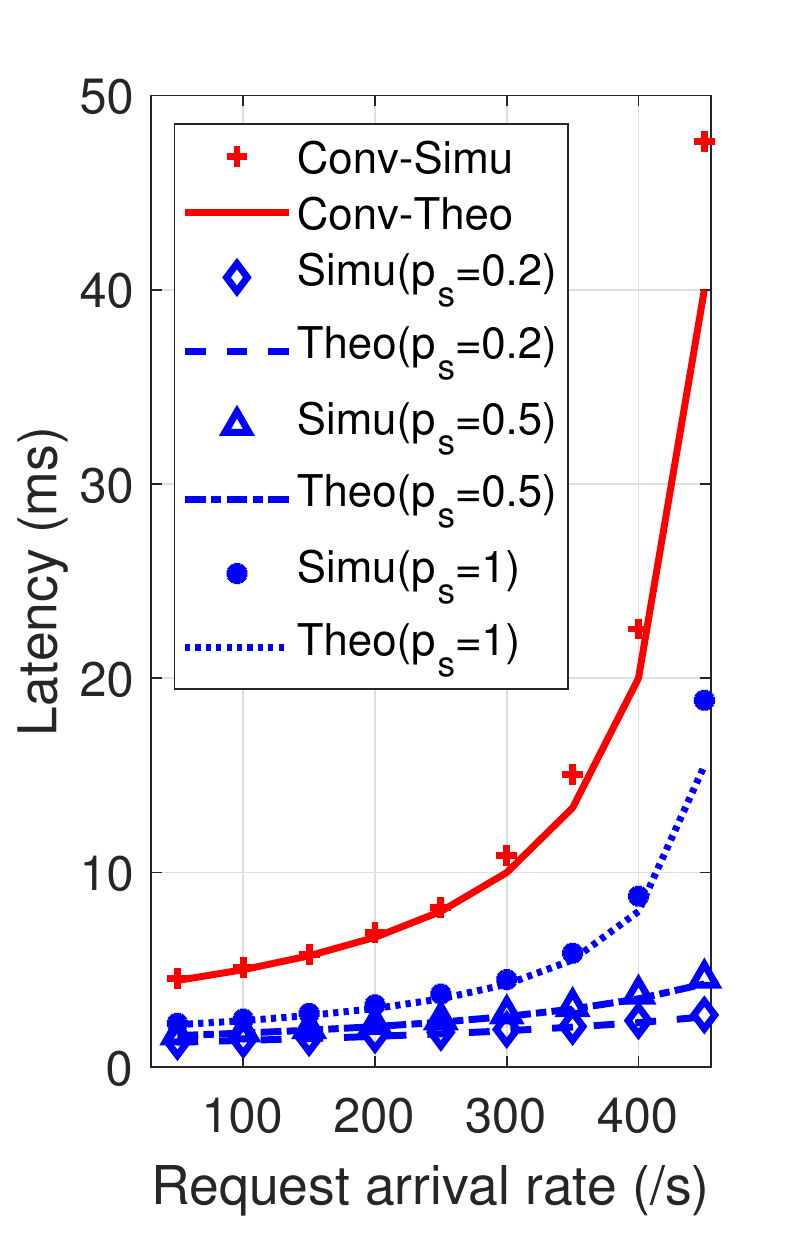}}
		%\hfil
		\subfloat[]{\includegraphics[width=1.6in]{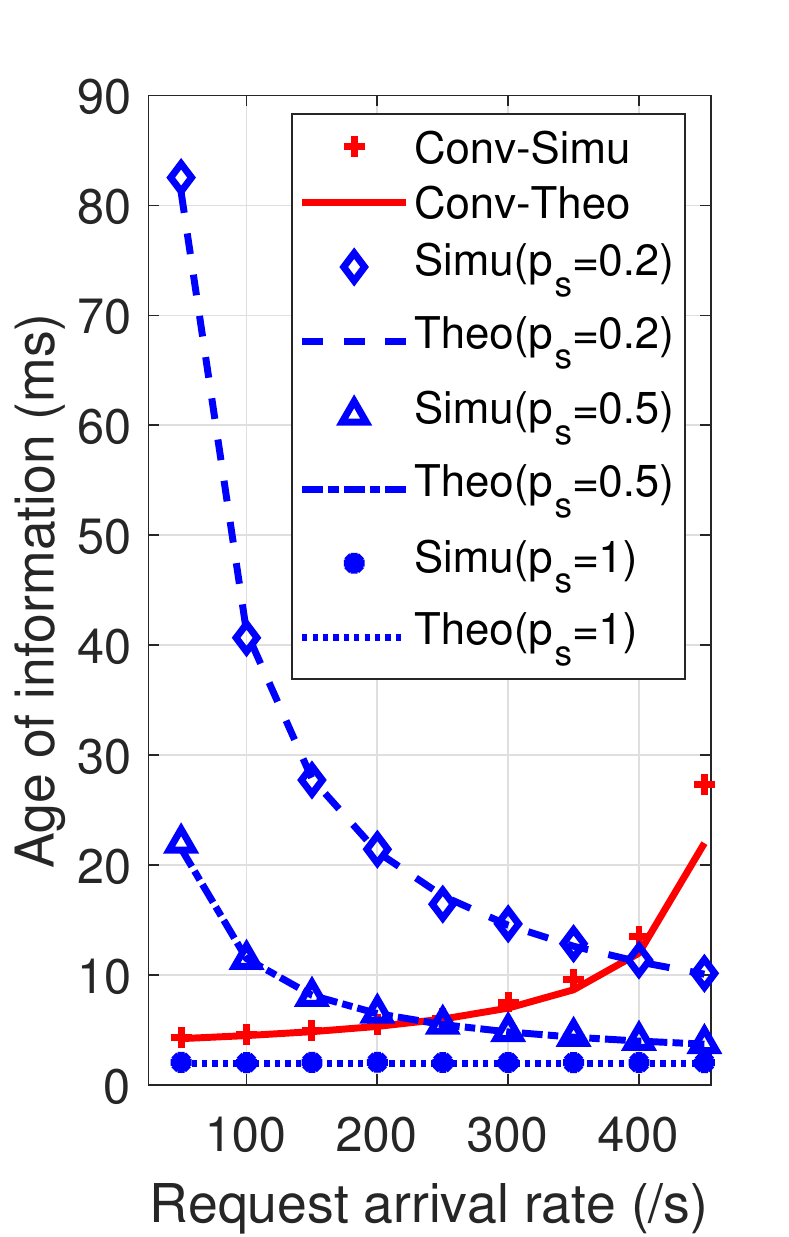}}
		%\hfil
		\caption{Analytical results validation of the ReA scheme, with conventional scheme as baseline (a) service latency, (b) age of information.}
		\label{fig_ReA_validation}
	\end{figure}
	
	This section conducts system-level simulations to validate the analytical results of AoI and service latency under the RSUC and ReA schemes, by implementing the OMNeT++ simulation platform.
	%The results also demonstrate the AoI-latency trade-off.
	In addition, the RSUC and ReA schemes are compared in terms of AoI and latency, under different system parameter settings.
	Furthermore, the service schemes are evaluated under a practical scenario, where SUMO simulates the real-trace vehicle mobility.
	
	%\begin{table}[!t]
	%	\caption{Simulation parameters}
	%	\label{tab_parameter}
	%	\centering
	%	\begin{tabular}{cccc}
	%		\hline
	%		\hline
	%		Parameter & Value & Parameter & Value \\
	%		\hline
	%		$L$ &  & $R$ &   \\
	%		$T$ & 1 ms &  & 1 Gb \\
	%		$\lambda_\mathrm{v}$ & 1.2 /s & $\lambda_\mathrm{p}$ & 4 /s \\
	%		$F$ & 1000 & $\nu$ & 0.56 \\
	%		$L_\mathrm{v}$ & 200 Gbps & $R_\mathrm{R}$ & 10 Gpbs \\ 	
	%		%$U_\mathrm{MBH}$ & 20 Mbps & $U_\mathrm{SBH}$ & 1 Mbps \\
	%		\hline
	%		\hline
	%	\end{tabular}
	%\end{table}

	%\begin{figure*}[!t]
	%	\centering
	%	\subfloat[] {\includegraphics[width=2.3in]{multi_cycle}}
	%	%\hfil
	%	\subfloat[]{\includegraphics[width=2.3in]{multi_ratio}}
	%	%\hfil
	%	\subfloat[]{\includegraphics[width=2.3in]{multi_latency}}
	%	\caption{Joint CALUD and resource allocation optimization with 5 source nodes, (a) optimal length of CALUD update cycle of each source, (b) ratio of radio resource allocated to each source, and (c) average service delay of each source.}
	%	\label{fig_multiple_node}
	%\end{figure*}
	
	\begin{figure}[!t]
		\centering
		{\includegraphics[width=2.5in]{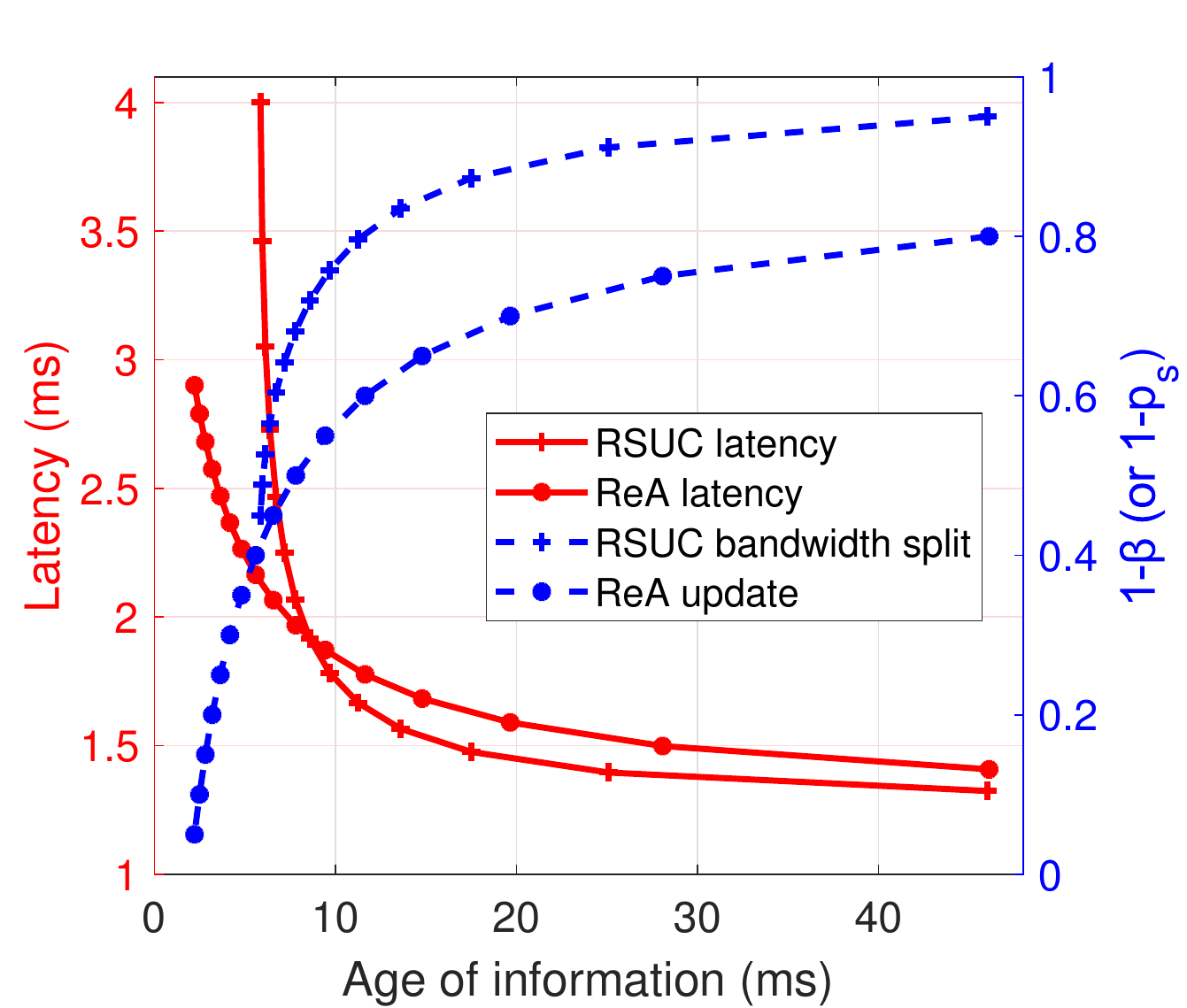}}
		\caption{AoI-Latency trade-off under the two schemes, content request arrival rate $\Lambda=200$ /s.}
		\label{fig_latency_AoI_tradeoff_single}
	\end{figure}
	
	\subsection{Analytical Results Validation}
	
	The analytical results of RSUC schemes are validated through simulations in case of a single publisher generating one item, as shown in Figs.~\ref{fig_RSUC_validation}.
	The content size $L$ is set to $3$ KB, the normalized service rate for cache update and content delivery are set to $R_\mathrm{UL}=1000$ and $R_\mathrm{DL}= 1000$ content items per second, respectively, while the content request arrival rate $\Lambda$ varies to reflect different traffic loads.
	Under the conventional scheme, the optimal bandwidth splitting is $\beta=0.5$, due to the symmetric channel gains of publishers and vehicles.
	The simulation results are obtained through OMNeT++ simulator, where the vehicle content requests and transmission time are generated randomly based on the Monte Carlo method.
	The analytical results are calculated based on Eqs.~(\ref{eq_delay_con_1}) and (\ref{eq_con_AoI}).	
	For the  RSUC scheme, the analytical results are based on the calculations of Eqs.~(\ref{eq_RC_delay}) and (\ref{eq_RC_AoI}), respectively, considering different bandwidth splitting ratio ($\beta$) settings for comparison.
	
	Figures~\ref{fig_RSUC_validation}(a) and \ref{fig_RSUC_validation}(b) show that the simulation results of both latency and AoI are quite close to the analytical ones under conventional and RSUC schemes, which validates the theoretical analysis.
	Under the conventional scheme, both the service latency and AoI increase with the content request arrival rate super-linearly.
	Under the RSUC scheme, the service latency increases while the AoI remains static as the request arrival rate varies for the given bandwidth splitting ratio $\beta$.
	This is reasonable since the RSUC scheme decouples the cache update and content delivery.
	In comparison, the RSUC scheme can perform better than the conventional scheme in terms of both latency and AoI when the network is heavily loaded (eg., $\beta=0.2$ or $\beta=0.5$).
	The reason is that RSUC reduces cache update frequency, and thus relieves traffic congestion.
	Furthermore, the service latency and AoI does not always trade off with respect to $\beta$.
	In specific, $\beta=0.5$ outperforms $\beta=0.8$ in both latency and AoI.
	This suggests that the bandwidth splitting ratio should not be too large, which is consistent with Theorem~2.
	
	% two groups of figures
	
	The analytical results of ReA scheme are also validated with different update probability, as shown in Fig.~\ref{fig_ReA_validation}.
	The analytical results of latency and AoI are calculated based on Eqs.~(\ref{eq_RA_delay}) and (\ref{eq_RA_AoI}), respectively.
	The simulation results are close to the analytical ones.
	Under the ReA scheme, the service latency increases with the request arrival rate, while the AoI decreases.
	Therefore, the ReA scheme is more advantageous in the case of heavy loads, compared with the conventional scheme.
	Furthermore, the service latency and AoI always shows a trade-off with respect to the cache update probability $p_s$.
	In specific, increasing $p_s$ results in higher latency but lower AoI, and $p_s=1$ can be treated as an extreme case.
	This result is consistent with Theorem 3.

	%	\begin{figure}[!t]
	%		\centering
	%		\subfloat[] {\includegraphics[width=1.6in]{Single_RSUC_beta2AOI}}
	%		%\hfil
	%		\subfloat[]{\includegraphics[width=1.6in]{Single_ReA_ps2AOI}}
	%		%\hfil
	%		\caption{Influences of operational parameters on the age of information (a) bandwidth splitting ratio of the RSUC scheme, (b) cache update probability under the ReA scheme.}
	%		\label{fig_tuning_parameter}
	%	\end{figure}

	\subsection{AoI and Latency Trade-off}
	
	% BS-centric

	Both the analytical and simulation results have demonstrated the existence of AoI-latency trade-off under the RSUC and ReA schemes if the cache update is restrained.
	For better understanding, we further study the AoI-latency trade-off, by tuning the bandwidth splitting and update probability under the two schemes, respectively.
	The AoI-latency trade-off is demonstrated by the solid lines in Fig.~\ref{fig_latency_AoI_tradeoff_single}, where the corresponding bandwidth splitting ratio $\beta$ and update probability $p_s$ are optimized based on problems (P3) and (P4) as the dash lines\footnote{$1-\beta$ and $1-p_s$ (instead of $\beta$ and $p_s$) are shown for better illustration, denoting the resource reserved for content delivery.}.
	The service latency is shown to first decrease and then level off as the AoI increases, under both schemes. 
	Meanwhile, both $\beta$ and $p_s$ decreases with AoI, indicating that less resources are needed for cache update if the freshness requirement is less strict.	
	The results of Fig.~\ref{fig_latency_AoI_tradeoff_single} reveal that we can reduce the service latency by sacrificing content freshness, which is especially significant when the AoI requirement is more strict.
	Regarding the AoI-latency trade-off, the ReA scheme is more beneficial when the AoI requirement is strict (i.e., less than 9 ms), and otherwise the RSUC scheme performs better.
	This result indicates that we need to choose the appropriate scheme based on the system parameters.	
	
	% request-adaptive
	
	\begin{figure}[!t]
		\centering
		{\includegraphics[width=2.5in]{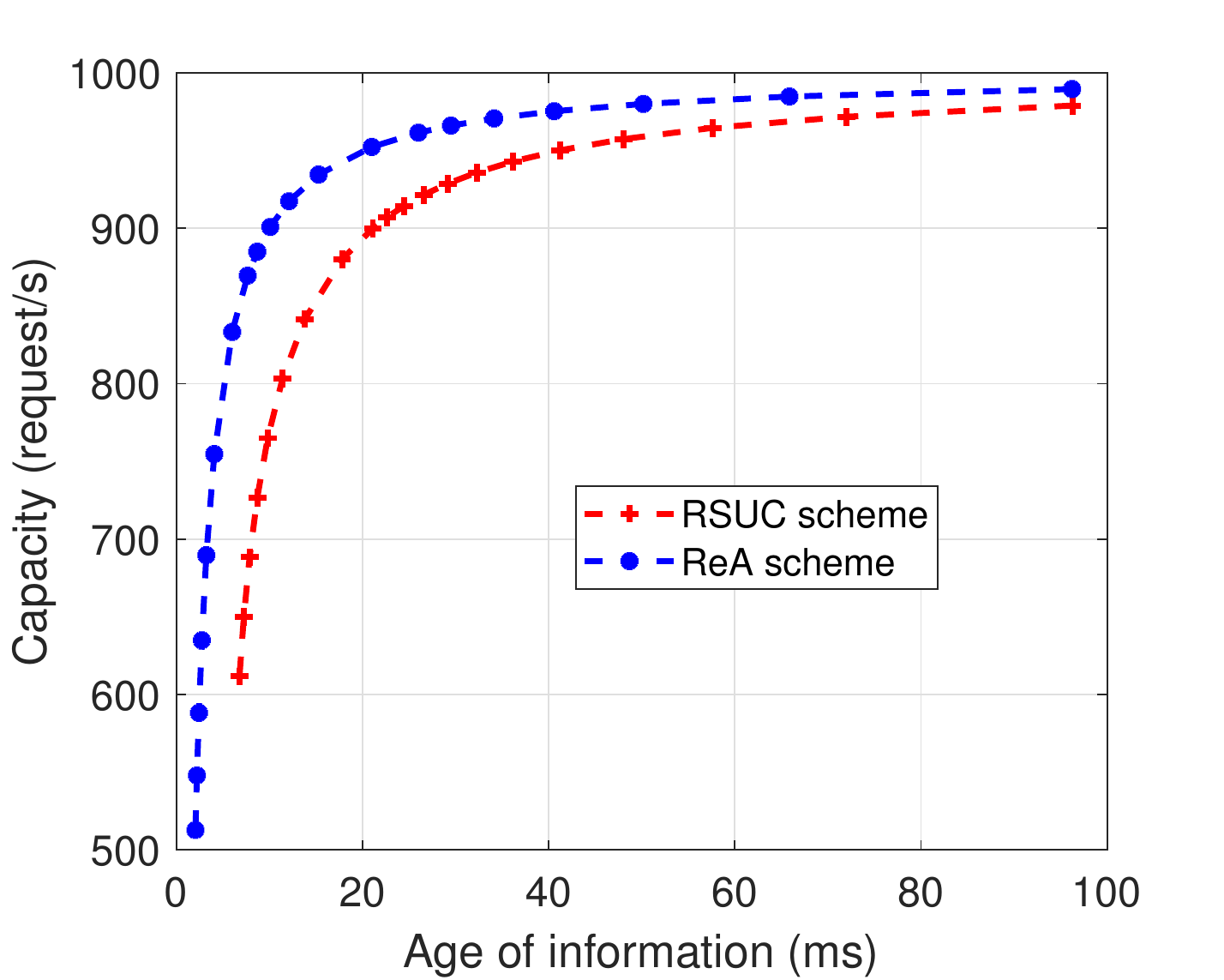}}
		\caption{Capacity-AoI trade-off under the two schemes.}
		\label{fig_Capacity_AoI_tradeoff_single}
	\end{figure}
	
	\begin{figure}[!t]
		\centering
		{\includegraphics[width=2.5in]{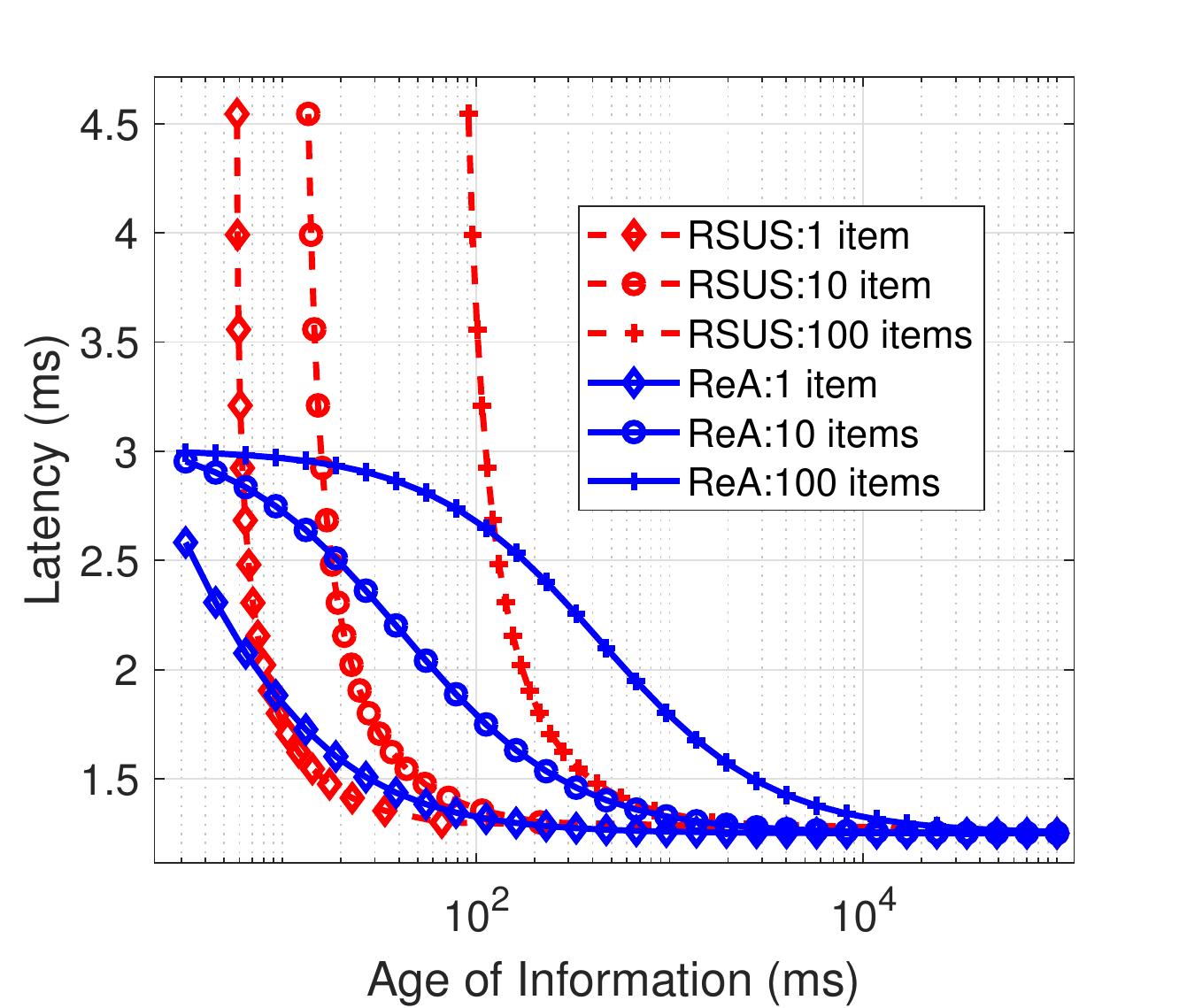}}
		\caption{Comparison of the two schemes, $R_\mathrm{UL}=R_\mathrm{DL}$=1000 /s, total content request arrival rate $\Lambda=200 $ /s, uniform item popularity.}
		\label{fig_comparison_multi_item_uniform}
	\end{figure}
	
	\begin{figure}[!t]
		\centering
		{\includegraphics[width=2.5in]{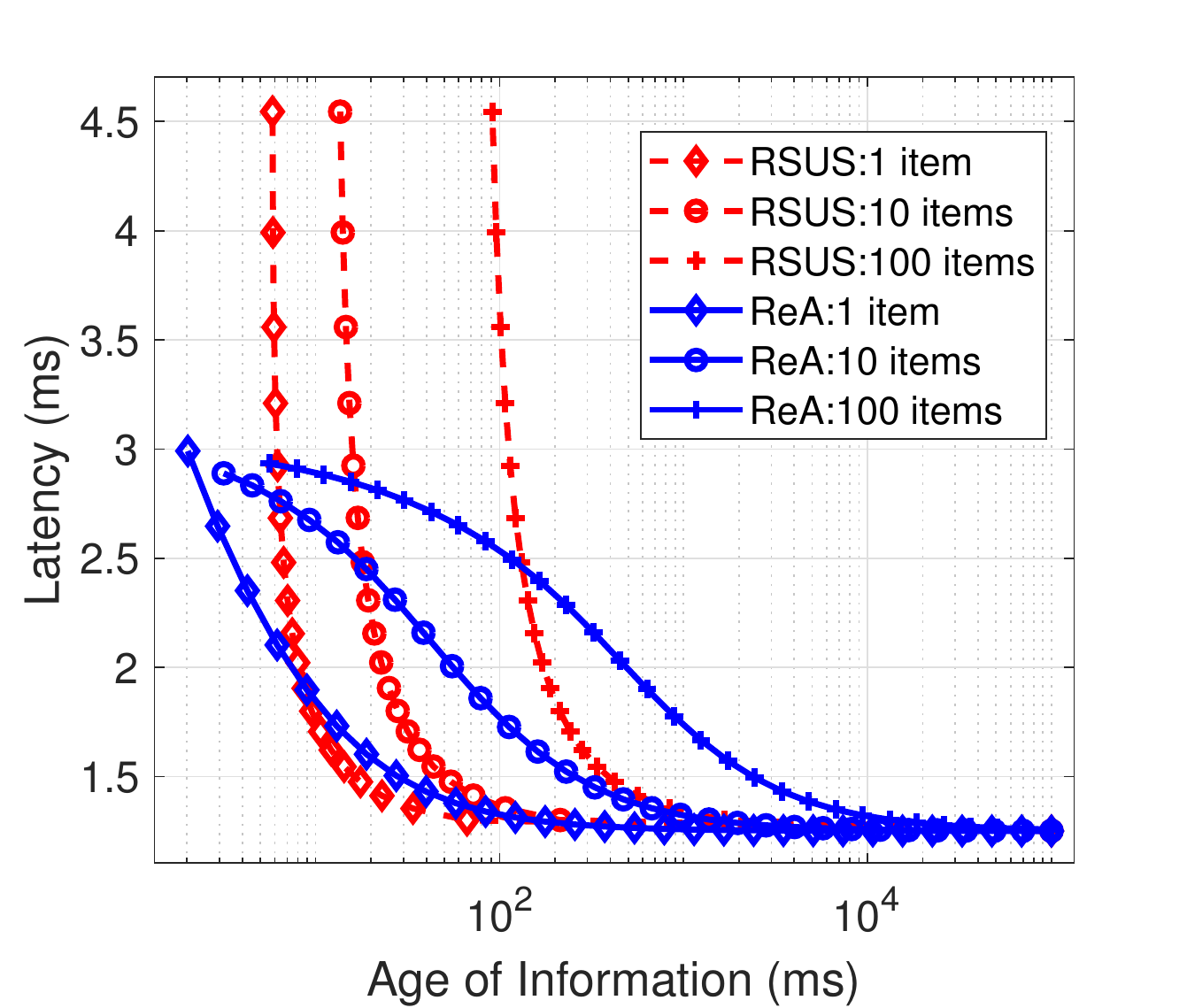}}
		\caption{Comparison of the two schemes,  $R_\mathrm{UL}=R_\mathrm{DL}$=1000 /s, total content request arrival rate $\Lambda=200$ /s, Zipf item popularity of exponent parameter 0.56.}
		\label{fig_comparison_multi_item_zipf}
	\end{figure}
	
	%With less strict AoI requirement, more content requests can be accommodated under the same service latency.
	Figure~\ref{fig_Capacity_AoI_tradeoff_single} shows the trade-off between AoI and capacity (i.e., the maximal request arrival rate that can be handled as the service latency goes to infinity).
	As AoI increases, the capacity first increases and then levels off under both schemes.
	In specific, the capacity converges to the normalized downlink service rate $R_\mathrm{DL}$, corresponding to $\beta=0$.
	Furthermore, the capacity increases significantly with AoI when the AoI requirement is strict, similar to the result of AoI-latency trade-off.

	\subsection{RSUC-ReA Comparison with Multiple Items}
	
	% different number of source nodes
	
	To offer insights into practical ICVN operations, we compare the performance of RSUC and ReA schemes considering the multi-publisher scenario, with respect to different system parameters.
	Fig.~\ref{fig_comparison_multi_item_uniform} shows the AoI-latency trade-off under the two schemes with different number of content items, where each item is requested with an equal probability.
	As the number of items increases, the trade-off curves of both schemes move rightwards, indicating performance degradation.
	The reason is that more resources are consumed to maintain the content freshness. 
	In comparison, the ReA scheme can achieve lower latency when the AoI requirement is smaller than a certain threshold.
	Furthermore, the ReA scheme is more advantageous as the number of items increases.

	The RSUC and ReA schemes are also compared considering different item popularity.
	In specific, the Zipf popularity distribution is considered, where the request probability of the $s$th most popular item is given by
	\begin{equation}
	p_s = \frac{1/s^{\theta}}{\sum_{i=1}^{S} (1/i^{\theta})},
	\end{equation}
	where the exponent $\theta$ reflects the concentration of requests.
	The typical value of $\theta$ is 0.56, corresponding to the video type services \cite{Gill07_youtube}.
	The comparison of the two schemes in case of Zipf-like popularity is shown in Fig.~\ref{fig_comparison_multi_item_zipf}.
	%The results are similar to that of the uniform item popularity.
	In specific, the performance of RSUC scheme remains the same whereas the performance of ReA scheme is slightly improved.
	This result is consistent with the analysis, where the RSUC scheme is not influenced by the request rate of individual items.
	
	%\subsection{Influence of Request Concentration}	
	
	\begin{figure}[!t]
		\centering
		{\includegraphics[width=2.5in]{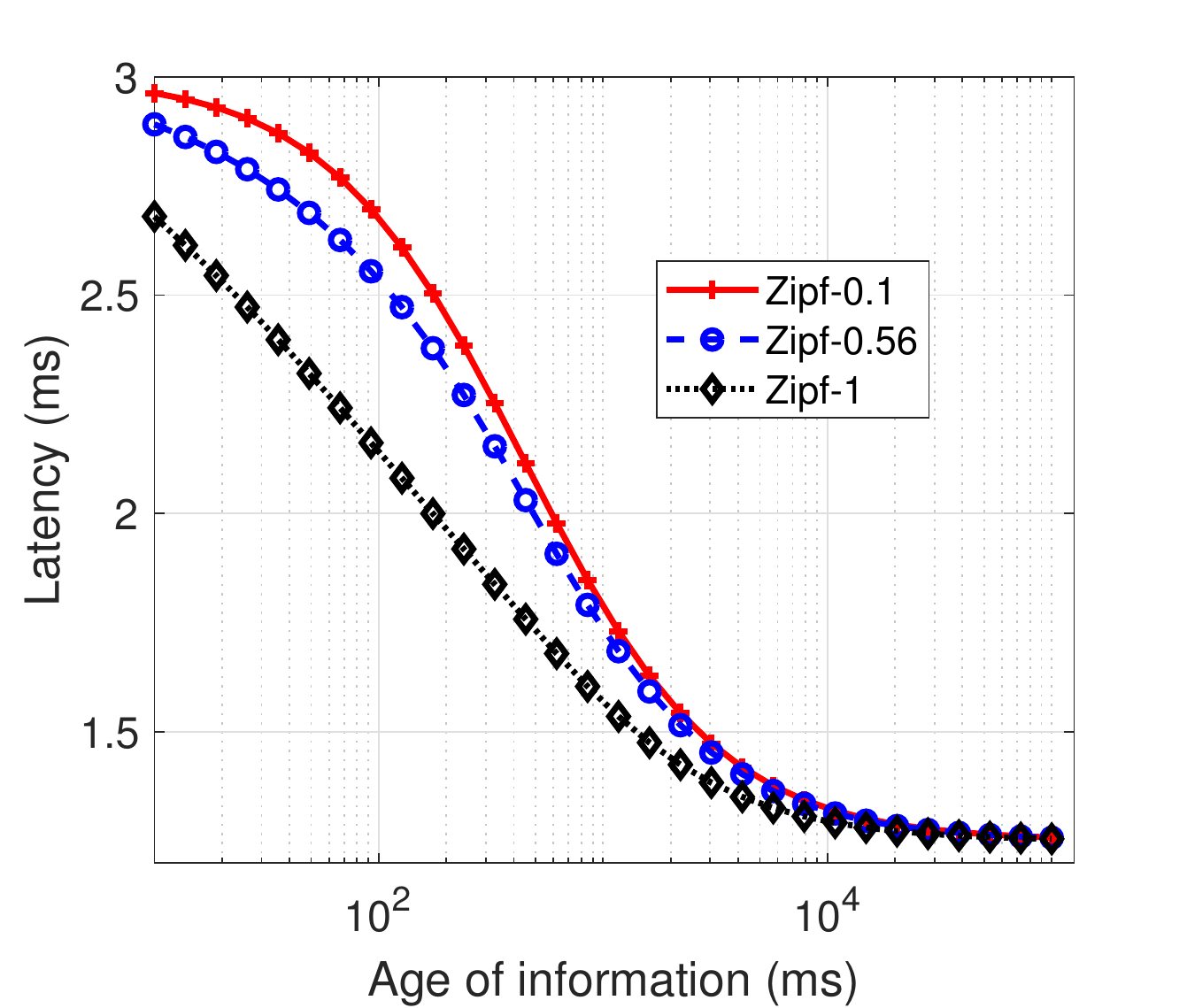}}
		\caption{Influence of content popularity (ReA scheme),  $R_\mathrm{UL}=R_\mathrm{DL}$=1000 /s, request arrival rate $\Lambda=200$ /s.}
		\label{fig_ReA_zipf}
	\end{figure}
	
	\begin{figure}[!t]
		\centering
		\subfloat[] {\includegraphics[width=1.6in]{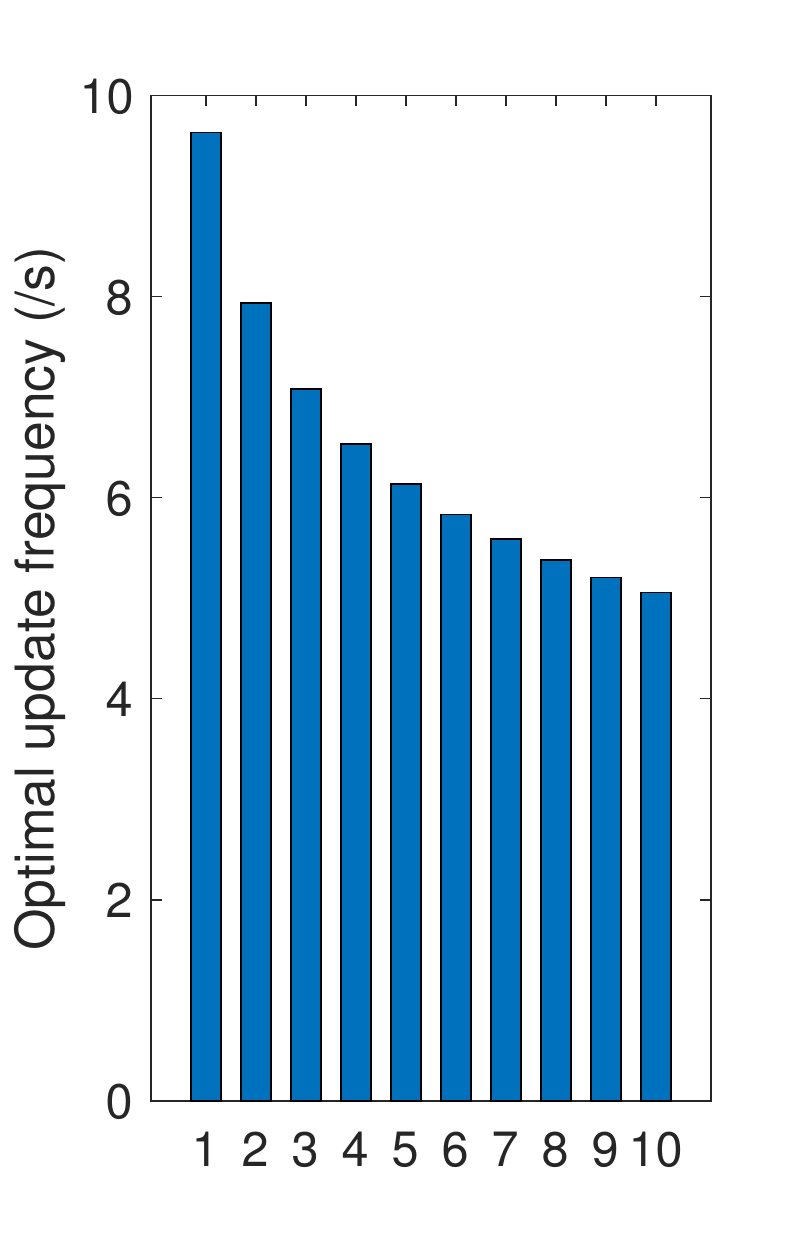}}
		%\hfil
		\subfloat[]{\includegraphics[width=1.6in]{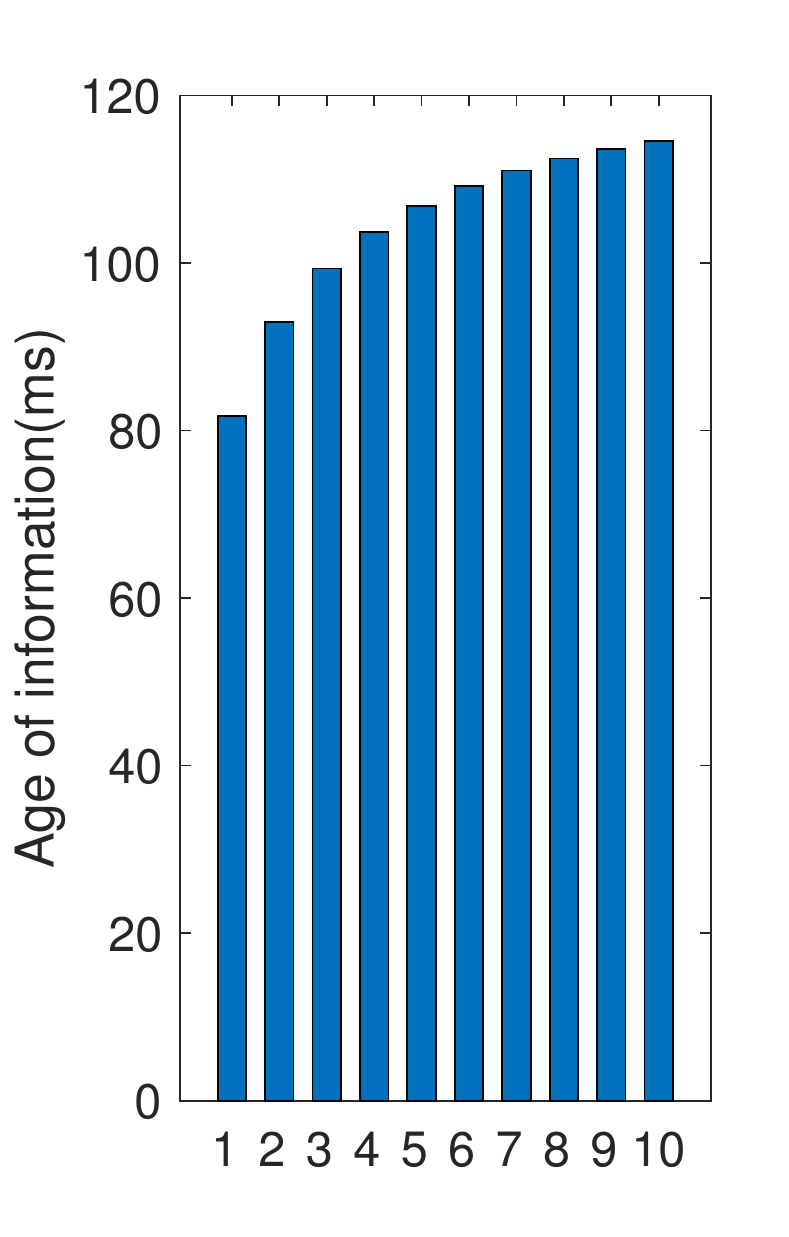}}
		%\hfil
		\caption{Optimal update control of individual items, (a) update probability, (b) age of information, average AoI 100 ms,  $R_\mathrm{UL}=R_\mathrm{DL}$=1000 /s, request arrival rate $\Lambda=200$ /s, Zipf exponent 0.56.}
		\label{fig_multiple_zipf}
	\end{figure}
	
	\begin{figure}[!t]
		\centering
		{\includegraphics[width=2.5in]{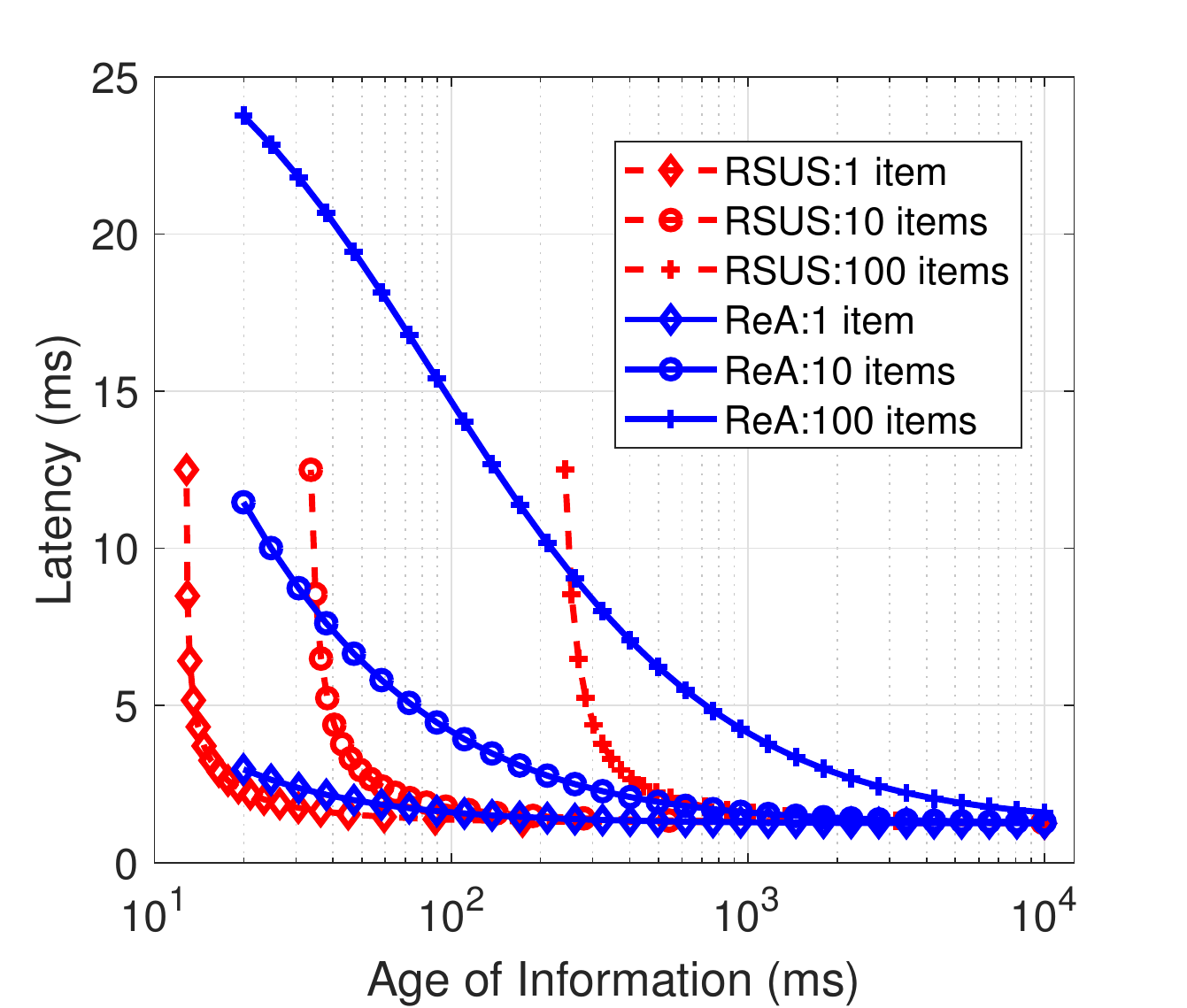}}
		\caption{Latency-AoI trade-off under the two schemes in case of asymmetric channels, $R_\mathrm{UL} = 300$ /s, $R_\mathrm{DL}=1000$ /s, request arrival rate $\Lambda=200$ /s, uniform content popularity.}
		\label{fig_latency_AoI_tradeoff_multiple_asys}
	\end{figure}

	The influence of request concentration on the ReA scheme is further illustrated, as shown in Fig.~\ref{fig_ReA_zipf}.
	The results demonstrate that the performance of the ReA scheme can be improved if the requests are more concentrated, i.e., larger Zipf exponent.
	Notice that the ReA scheme can adjust update frequency of each item, based on the corresponding arrival rate.
	This avoids frequent update of unpopular contents, and improves efficiency.
	Fig.~\ref{fig_multiple_zipf} further shows the update frequency and AoI of each item, under the ReA scheme.
	10 items are considered with Zipf popularity of 0.56, and the average AoI requirement is 100 ms.
	The update probability of each item is different, where the less popular contents are updated less frequently.
	Accordingly, the AoI of less popular contents are higher.
	This result is consistent with the analysis in \cite{Yates17_AoI_cache_update_17}, where the less popular contents are set with a lower weight when calculating the average AoI.
	The results of Figs.~\ref{fig_ReA_zipf} and \ref{fig_multiple_zipf} reveal the advantages of the ReA scheme when dealing with heterogeneous content requests.

	\begin{figure}[!t]
		\centering
		{\includegraphics[width=2.5in]{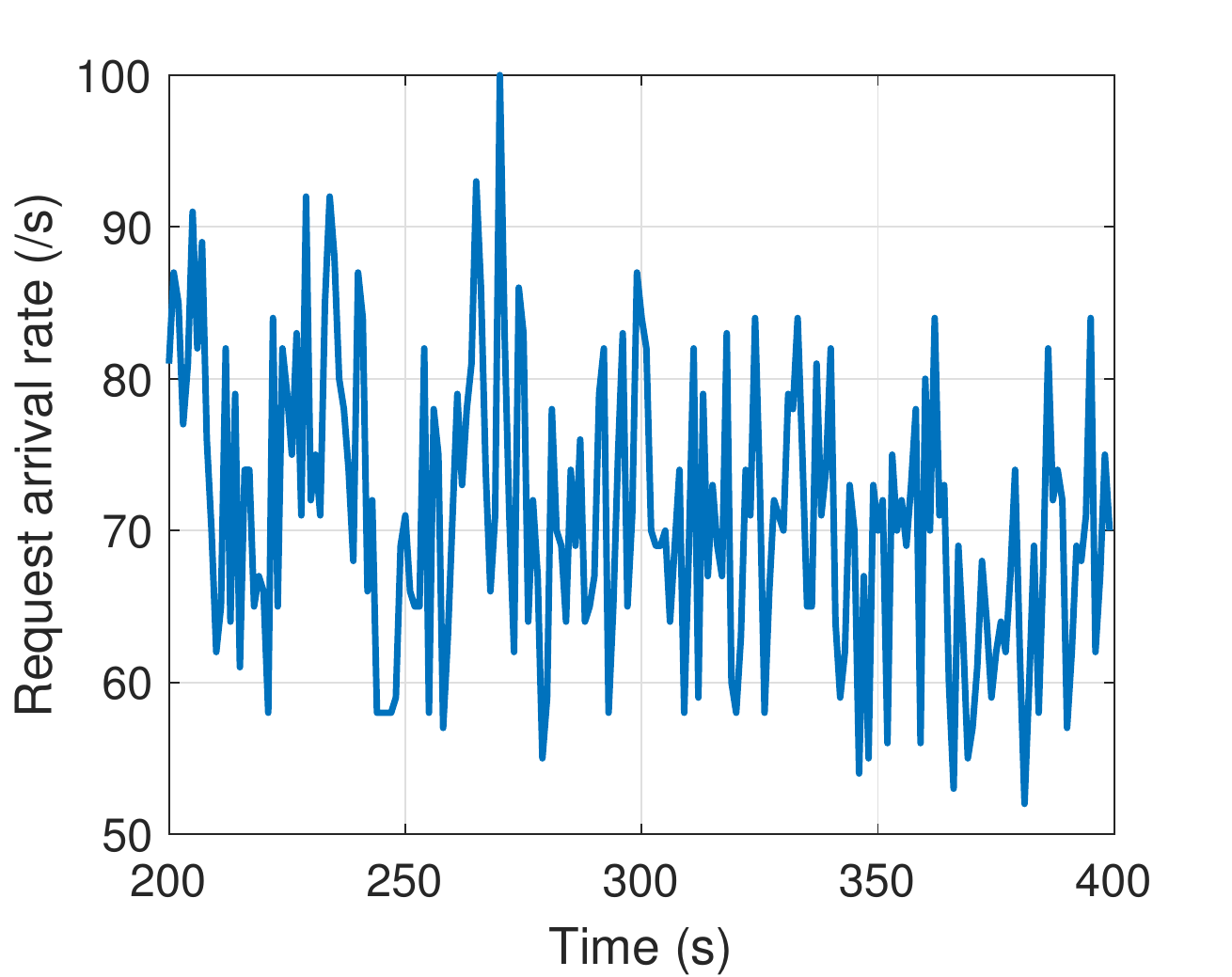}}
		\caption{Illustration of time-varying request arrival rate at the RSU, vehicle mobility trace generated by SUMO.}
		\label{fig_RT_lambda_time}
	\end{figure}
	
	\begin{figure*}[!t]
		\centering
		\subfloat[] {\includegraphics[width=2.5in]{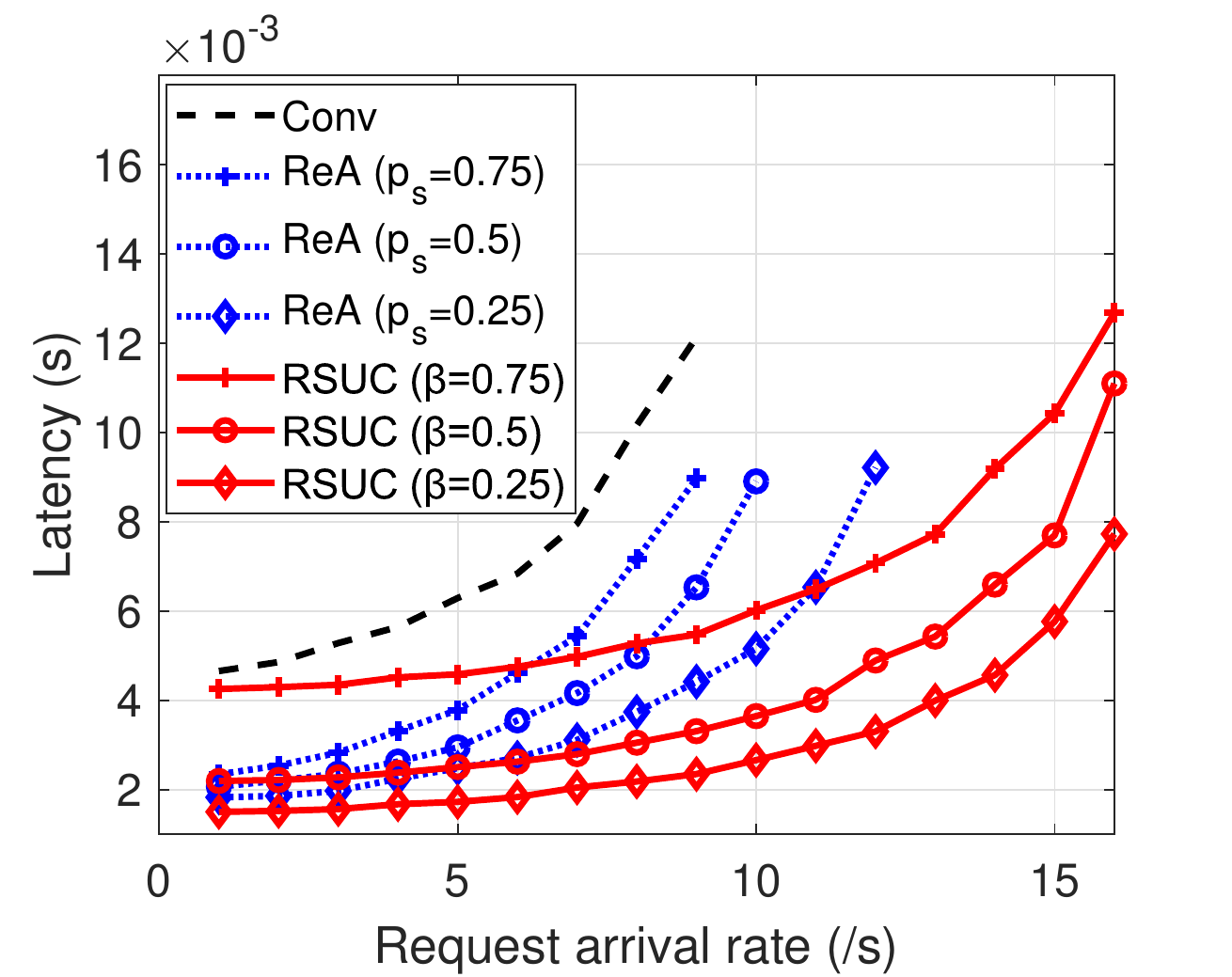}}
		%\hfil
		\subfloat[]{\includegraphics[width=2.5in]{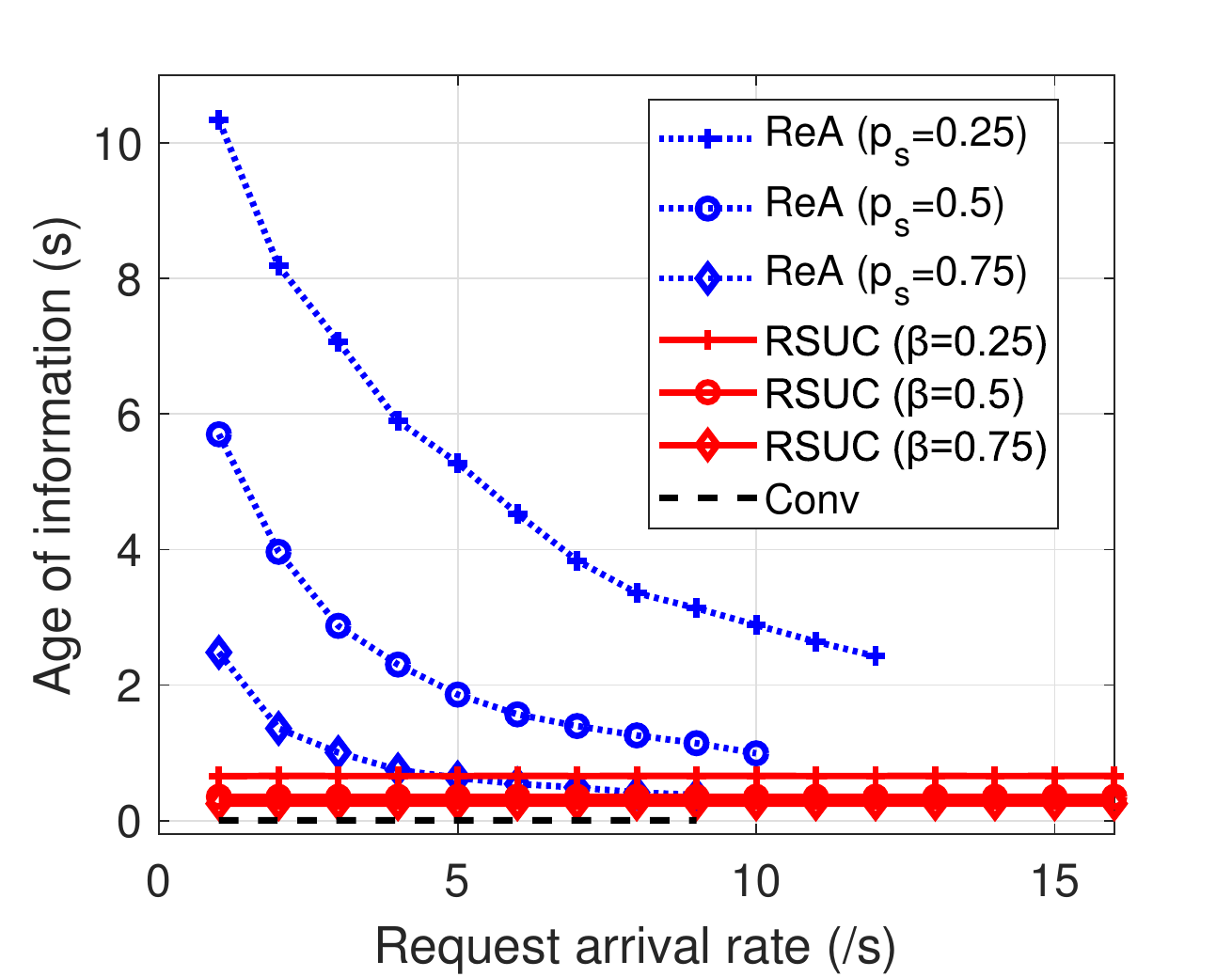}}
		%\hfil
		\caption{Performance evaluation of three schemes based on SUMO and OMNeT++ experiments.}
		\label{fig_RT_experiment}
	\end{figure*}
	
	The influence of asymmetric channels are also investigated.
	In practice, the channel transmission rates from publishers to RSUs may be much lower than that from RSUs to vehicles, considering the equipped hardware like transceivers and antennas.
	By setting $R_\mathrm{UL} = 300$ /s, $R_\mathrm{DL}=1000$ /s, we evaluate performance of the two schemes, as shown in Fig.~\ref{fig_latency_AoI_tradeoff_multiple_asys}.
	The performance of both schemes degrades significantly, compared with the case of $R_\mathrm{UL}=R_\mathrm{DL}$=1000 /s in Fig.~\ref{fig_comparison_multi_item_uniform}.
	Notice that the performance of the RSUC scheme degrades more with larger number of items, and the minimal AoI achieved can increase very fast.
	By contrast, the minimal AoI achieved by the ReA scheme is not influenced by the number of items, whereas the corresponding latency increases.
	However, the RUSC scheme completely outperforms the ReA scheme in case of a single item, as shown by the lines with rhombuses.

	Based on the comparison of the two schemes, we come to following conclusions: 	
	(1) The ReA scheme performs better than the RSUC scheme if the average AoI requirement is smaller than a certain threshold, and vice versa; 	
	(2) The ReA scheme is more advantageous with a larger number of items; 	
	(3) The ReA scheme adjusts update frequency of individual items based on the popularity, which is more suitable for the heterogeneous application scenarios.

	\subsection{Real-trace Simulation}

	To evaluate the performance of the proposed RSUC and ReA schemes in practical scenarios, we build a network-level simulation platform based on SUMO and OMNeT++ simulators.
	We consider a RSU deployed at an intersection in the city of Erlangen, Germany, which represents a typical urban scenario. 
	100 source nodes are deployed, which are randomly located within a range of 200 m to the RSU.
	Each source node generates one content item, and the content items are sent to the RSU according to the implemented cache update scheme.	
	The mobility traces of vehicles are generated by the SUMO simulator.
	Each vehicle raises requests randomly at a certain rate, and all source share equivalent popularity.
	Accordingly, the requests arrival rate at the RSU varies randomly with vehicle mobility, as illustrated in Fig.~\ref{fig_RT_lambda_time}.
	The packet size is set to 3 KB, and the transmission rates are set to 24 Mbps for cache update and content delivery, respectively, according to the 802.11p standard\footnote{Different from the analytical model, the transmission rates of uplink and downlink are both constant on the OMNeT++ simulator.}.
	
	We implement the proposed RSUC and ReA schemes at the RSU, and the conventional scheme is also conducted as a baseline.
	The service latency and AoI of each request is recorded, whereby the average value is calculated.
	Figure~\ref{fig_RT_experiment} shows the results of different schemes with specified parameter settings, where the x-axis represents the request arrival rate per vehicle.
	The results show that the conventional scheme achieves the minimal average AoI while the service latency is the highest.
	Furthermore, the service capacity of the conventional scheme is also the lowest, which is consistent with the analytical results.
	For example, the network becomes overloaded as the request arrival rate approaches 12 /s under the conventional scheme.
	
	The results demonstrate that both the ReA and RSUC schemes can balance latency and AoI by adjusting the corresponding operational parameters. 
	In comparison, the two schemes can significantly reduce the average latency when the traffic load is high.
	When the traffic arrival rate is 9 /s, the RSUC scheme with bandwidth splitting ratio of 0.25 can reduce the average latency by nearly 80\% compared with the conventional scheme, while the ReA scheme can achieve 65\% reduction.
	The average AoI will increase to around 0.2 s, which is acceptable for applications whose contents varies at second- or minute-levels. 
	This result indicates the effectiveness to leverage mobile edge caching in ICVNs, while appropriate cache update management schemes is needed.
	
	In general, the RSUC scheme outperforms the ReA scheme according to the simulations.
	This reason is that the popularity of each source is uniform, and thus the ReA scheme cannot live up to the full potential of fine-grained update frequency.
	In addition, the update cycle of ReA scheme shows uncertain depending on the random arrival of requests, which may also degrade the freshness performance on average.
	However, the ReA scheme may outperform the RSUC scheme in case of concentrated requests, which will be studied in future work.

%%%%%%%%%%%%%%%%%%%%%%%%%%%%%%%%%%%%%%%%%%%%%%%%%%%%%%%%%%%%%%%%%%%%%%%%%%%%%%%%%%%%%%%%%%%%%%%%%%%
\section{Conclusions and Future Work}
\label{sec_conclusions}

The RSUC and ReA schemes have been proposed to support the dynamic driving-related context information service under the ICVN architecture, through the joint design of cache update and content delivery. 
Under the proposed schemes, the interplay between service latency and content freshness has been studied.
Analytical results have shown that the service latency and content freshness demonstrate a trade-off when the cache update frequency is restrained, but can degrade simultaneously if the RSU cache is frequently updated.  
In this regard, the RSUC and ReA schemes have been optimized to balance service latency and content freshness on demand.
Simulation results have shown that both the RSUC and ReA schemes can effectively improve content freshness and service latency, compared with the conventional freshness-first scheme.
Furthermore, the ReA scheme outperforms the RSUC scheme in case of strict freshness requirements, heavy traffic loads, larger number of content items, or concentrated content interests.
Future studies will further consider the inequivalent content sizes and the co-existence of differentiated content types. 

%%%%%%%%%%%%%%%%%%%%%%%%%%%%%%%%%%%%%%%%%%%%%%%%%%%%%%%%%%%%%%%%%%%%%%%%%%%%%%%%%%%%%%%%%%%%%%%%%%%

%\appendices
%   \input{appendix.tex}
   
%   % use section* for acknowledgment
%   \ifCLASSOPTIONcompsoc
%   % The Computer Society usually uses the plural form
%   \section*{Acknowledgments}
%   \else
%   % regular IEEE prefers the singular form
%   \fi

\ifCLASSOPTIONcaptionsoff
   \newpage
   \fi

\normalem


\end{document}